# POR for Security Protocol Equivalences[⋆]
## Beyond Action-Determinism

David Baelde[1], Stéphanie Delaune[2], and Lucca Hirschi[3]

[1] LSV, ENS Paris-Saclay & CNRS, Inria Paris, Université Paris-Saclay, France
[2] Univ Rennes, CNRS, IRISA, France
[3] Department of Computer Science, ETH Zurich, Switzerland

**Abstract.** Formal methods have proved effective to automatically analyze protocols. Over the past years, much research has focused on verifying *trace equivalence* on protocols, which is notably used to model many interesting *privacy* properties, *e.g.*, anonymity or unlinkability. Many tools for checking trace equivalence rely on a naive and expensive exploration of all interleavings of concurrent actions, which calls for partial-order reduction (POR) techniques. In this paper, we present the first POR technique for protocol equivalences that does not rely on an action-determinism assumption: we recast the trace equivalence problem as a reachability problem, to which persistent and sleep set techniques can be applied, and we show how to effectively apply these results in the context of symbolic executions. We report on a prototype implementation, improving the tool DeepSec.

## 1 Introduction

Security protocols are notoriously difficult to design, and their flaws can have a huge impact. Leaving aside implementation flaws and weaknesses of cryptographic primitives, there is already a long history of logical mistakes in the basic design of protocols, *e.g.*, [28,6,12,5]. At this level of detail, protocols can however be represented in the so-called symbolic model, which makes them amenable to automated formal verification. This approach has lead to mature tools and industrial successes, *e.g.*, [13,7,29].

Verification techniques have focused at first on reachability properties of protocols, used to model, *e.g.*, the secrecy or authenticity or messages. More recently, equivalence properties have received a lot of attention, as they are often necessary to model privacy properties such as ballot secrecy in e-voting [23], anonymity or unlinkability [5,14]. Equivalence verification is complex, and the various state-of-the-art techniques each have their own limitations. Tools for verifying scenarios with an unbounded number of sessions such as ProVerif [13] or Tamarin [32] are usually efficient but only support a constrained form of equivalence, namely *diff-equivalence*, which is too limiting, *e.g.*, to model unlinkability [27]. Many tools for verifying bounded scenarios rely on symbolic execution [22]. For instance, Apte [16] and its successor DeepSec [19] implement an

[⋆] This work has been partially supported by the European Research Council (ERC) under the European Union's Horizon 2020 research and innovation program (grant agreement No 714955-POPSTAR), as well as from the French National Research Agency (ANR) under the project TECAP.

algorithm that explores all symbolic executions, maintaining pairs of sets of symbolic states and solving, at each step, complex equality, deducibility and indistinguishability constraints. Akiss [15] follows a different approach, enumerating all symbolic executions to check that none yields an non-equivalence witness. The strength of these tools is that they decide *trace equivalence*, which can adequately capture privacy goals such as unlinkability. However, their algorithms are very costly. Although much progress has been made, it is still only possible to analyze relatively small scenarios in a reasonable amount of time.

All the techniques mentioned above for deciding trace equivalence of security protocols rely on an enumeration of all symbolic executions including all interleavings of concurrent actions. This is obviously a cause of major inefficiency, which has lead to the quest for *partial-order reduction* (POR) techniques. These techniques, which have a long and successful history in traditional software verification [26,31,10], generally consist in leveraging action independencies to restrict the interleavings that a model-checking algorithm explores. In the context of verifying reachability properties for security protocols, some specific POR techniques have sometimes had to be devised [20,30], but there have also been successful uses of generic POR techniques such as *sleep sets* [21] (see [9] for a detailed discussion). In the context of verifying *equivalences* for security protocols, the only available POR techniques are, to the best of our knowledge, the ones we proposed in [8,9]. These *ad hoc* techniques have lead to significant performance gains in Apte and DeepSec [19,9]. However, they crucially rely on an *action-determinism* assumption (*i.e.*, once the observable trace is fixed, the system is deterministic) which is very limiting in practice. For instance, there is no precise modeling of unlinkability involving action-deterministic systems.

In this paper, we present the first POR technique for checking trace equivalence on security protocols, without any action-determinism assumption. Our first step towards this goal is to recast the trace equivalence problem as a reachability problem in a carefully designed labeled transition system (LTS), to which we can then apply persistent and sleep set techniques. However, this result is not directly useful in practice, for several reasons. First, this LTS is infinitely branching, due to the arbitrary choices that the attacker can make when interacting with the protocol. This is the main issue addressed in protocol equivalence checkers, typically through symbolic execution. Second, determining when two actions are independent (the first ingredient of POR techniques) is far from obvious in our LTS. Independencies are often approximated through simple static checks in practical POR algorithms [26] but, as we shall see, it does not seem feasible in our setting without losing too many independencies. Instead, we determine independencies by exploring symbolic executions. We ignore constraint solving in that process, as it would be too expensive: this trade-off allows us to detect enough independencies at a reasonable cost. More generally, we show how to compute persistent sets in the same style, to eventually obtain a symbolic form of the sleep set technique. Third, the direct symbolic approach would still be overly expensive, due to another typical state explosion problem caused by conditionals [11]. We circumvent it by showing that conditionals can be simplified, and often eliminated, in a way that does not affect persistent set computations. This approach yields a POR technique that is fast enough and allows to significantly reduce the number of symbolic traces to consider when checking equiv-alences. The technique is also independent of the specific verification algorithms that will be used to check equivalence along the reduced set of traces. We implement it as a library, and integrate it in DeepSec to validate the approach experimentally.

*Outline.* We present a standard security protocol model in Section 2. After recalling persistent and sleep set techniques in Section 3, we design in Section 4 our concrete equivalence LTS to which they apply. Section 5 then defines a symbolic abstraction of this LTS, and shows how it can be used to obtain effective POR algorithms, notably through the collapse of conditionals. Finally, we present our implementations and experimental results in Section 6. Detailed proofs and additional examples are given in Appendices A,B,C,D.

## 2 Model for security protocols

As is common, we model security protocols in a variant of the applied pi-calculus [2]: processes exchange messages represented by terms quotiented by an equational theory.

### 2.1 Syntax

We assume a number of disjoint and infinite sets: a set $\mathcal{C}h$ of channels, denoted by $c$ or $d$; a set $\mathcal{N}$ of *names*, denoted by $n$ or $k$; a set $\mathcal{X}$ of *variables*, denoted by $x$, $y$, $z$; and a set $\mathcal{W}$ of *handles* of the form $\mathsf{w}_{c,i}$ with $c \in \mathcal{C}h$ and $i \in \mathbb{N}$, which will be used for referring to previously output terms. Next, we consider a *signature* $\Sigma$ consisting of a set of function symbols together with their arity. Terms over a set of atomic data $A$, written $\mathcal{T}(A)$, are inductively generated from $A$ and function symbols from $\Sigma$. When $A \subseteq \mathcal{N}$, elements of $\mathcal{T}(A)$ are called *messages* and written $m$. When $A \subseteq \mathcal{W}$, they are called *recipes* and written $M$, $N$. Intuitively, recipes express how a message has been derived by the environment from the messages obtained so far. Finally, we consider an equational theory $\mathsf{E}$ over terms to assign a meaning to function symbols in $\Sigma$.

Protocols are then modelled through *processes* using the following grammar:

$P, Q := 0 \mid \mathtt{in}(c, x).P \mid \mathtt{out}(c, u).P \mid \mathtt{if}\ u = v\ \mathtt{then}\ P\ \mathtt{else}\ Q \mid (P \mid Q) \mid P + Q$

where $c \in \mathcal{C}h$, $u, v \in \mathcal{T}(\mathcal{N} \uplus \mathcal{X})$ and $x \in \mathcal{X}$. The process $0$ does nothing. The process $\mathtt{in}(c, x).P$ expects a message $m$ on the public channel $c$, and then behaves like $P\{x \mapsto m\}$, *i.e.*, $P$ in which $x$ has been replaced by $m$. The process $\mathtt{out}(c, u).P$ outputs $u$ on the public channel $c$, and then behaves like $P$. We have constructions to perform tests (modulo $\mathsf{E}$), parallel composition, and non-deterministic choice. We do not consider replication, and thus we do not need a specific "new" operation: we assume that names are implicitly freshly generated.

*Example 1.* Consider the signature $\Sigma = \{\mathsf{enc}, \mathsf{dec}, \mathsf{h}\}$, where $\mathsf{h}$ is a free symbol representing an hash function, whereas symmetric encryption will be modelled through the equation $\mathsf{dec}(\mathsf{enc}(x, y), y) = x$. We model a simple challenge/response protocol where an agent sends a fresh random number $n$ encrypted with a shared symmetric key $k$ as a challenge, then waits for an answer of the form $\mathsf{enc}(\mathsf{h}(n), k)$, as follows:

$P_{\mathsf{challenge}} = \mathtt{out}(c, \mathsf{enc}(n, k)).\mathtt{in}(c, x).\mathtt{if}\ x = \mathsf{enc}(\mathsf{h}(n), k)\ \mathtt{then}\ P_{\mathsf{success}}\ \mathtt{else}\ P_{\mathsf{fail}}$
$\quad \mid \mathtt{in}(d, y).\mathtt{out}(d, \mathsf{enc}(\mathsf{h}(\mathsf{dec}(y, k)), k)).0$

### 2.2 Semantics

A configuration $\mathcal{K}$ is a pair $(\mathcal{P}; \phi)$ such that:
- $\mathcal{P}$ is a multiset of processes with no free variable, or a special object $\bot_i$ with $i \in \mathbb{N}$;
- $\phi = \{w_1 \triangleright m_1, \ldots, w_n \triangleright m_n\}$ is a *frame*, *i.e.*, a substitution where $w_1, \ldots, w_n$ are handles in $\mathcal{W}$, and $m_1, \ldots, m_n$ are messages, *i.e.*, terms in $\mathcal{T}(\mathcal{N})$.

Configurations $(\bot_i, \phi)$ are called *ghost configurations* dead at age $i$, and will only become useful in Section 4. Other configurations are said to be *alive*.

Operational semantics are given as an LTS on (alive) configurations, with the relation $\xmapsto{\alpha}$ defined next. There, the index of the next output to be performed on channel $c$ is defined as $\#_c(\mathrm{dom}(\phi)) = \max(\{0\} \cup \{j+1 \mid w_{c,j} \in \mathrm{dom}(\phi)\})$.

$(\{\mathtt{in}(c,x).Q\} \uplus \mathcal{P}; \phi) \xmapsto{\mathtt{in}(c,M)} (\{Q\{x \mapsto M\phi\}\} \uplus \mathcal{P}; \phi) \qquad \text{if } M \in \mathcal{T}(\mathrm{dom}(\phi))$

$(\{\mathtt{out}(c,u).Q\} \uplus \mathcal{P}; \phi) \xmapsto{\mathtt{out}(c,w_{c,i})} (\{Q\} \uplus \mathcal{P}; \phi \cup \{w_{c,i} \triangleright u\}) \quad \text{with } i = \#_c(\mathrm{dom}(\phi))$

$(\{\mathtt{if}\ u = v\ \mathtt{then}\ Q_1\ \mathtt{else}\ Q_2\} \uplus \mathcal{P}; \phi) \xmapsto{\tau} (\{Q_1\} \uplus \mathcal{P}; \phi) \qquad \text{if } u =_{\mathsf{E}} v$

$(\{\mathtt{if}\ u = v\ \mathtt{then}\ Q_1\ \mathtt{else}\ Q_2\} \uplus \mathcal{P}; \phi) \xmapsto{\tau} (\{Q_2\} \uplus \mathcal{P}; \phi) \qquad \text{if } u \neq_{\mathsf{E}} v$

$(\{Q_1 + Q_2\} \uplus \mathcal{P}; \phi) \xmapsto{\tau} (\{Q_1\} \uplus \mathcal{P}; \phi) \qquad (\{Q_1 + Q_2\} \uplus \mathcal{P}; \phi) \xmapsto{\tau} (\{Q_2\} \uplus \mathcal{P}; \phi)$

$(\{Q_1 \mid Q_2\} \uplus \mathcal{P}; \phi) \xmapsto{\tau} (\{Q_1, Q_2\} \uplus \mathcal{P}; \phi) \quad (\{0\} \uplus \mathcal{P}; \phi) \xmapsto{\tau} (\mathcal{P}; \phi)$

A process may input a term that an attacker built using the knowledge available to him through the frame, where messages output by the protocol are added. The output rule slightly differs from the standard one, which would use a fresh handle variable. Our use of fixed constants $w_{c,i}$ makes it possible to view the transition system as a standard LTS, without any notion of freshness or $\alpha$-renaming. Anticipating on the next sections where we build on top of this a different LTS encoding trace equivalence, we note that this design choice does not create spurious dependencies. We do not need to model internal communications since we assume public channels: all communications are controlled by the attacker. The last rules evaluate conditionals (modulo E), break parallel operators, remove null processes, and perform non-deterministic choices.

The relation $\mathcal{K}_0 \xmapsto{\alpha_1 \ldots \alpha_k} \mathcal{K}$ between configurations, where $k \geq 0$ and each $\alpha_i$ is an observable or a $\tau$ action, is defined in the usual way. Given a sequence tr of actions, we denote obs(tr) the sequence of actions obtained by erasing $\tau$ actions.

*Example 2.* Continuing Example 1, and starting with $\mathcal{K}_0 = (P_{\mathsf{challenge}}; \emptyset)$ we have that:

$$\mathcal{K}_0 \xmapsto{\mathtt{out}(c,w_{c,0}).\mathtt{in}(d,w_{c,0}).\mathtt{out}(d,w_{d,0}).\mathtt{in}(c,w_{d,0}).\tau} (P_{\mathsf{success}}; \phi)$$

where $\phi =_{\mathsf{E}} \{w_{c,0} \triangleright \mathsf{enc}(n,k),\ w_{d,0} \triangleright \mathsf{enc}(\mathsf{h}(n),k)\}$.

### 2.3 Equivalences

Many privacy-type properties (*e.g.*, ballot secrecy in evoting, unlinkability) are modelled relying on trace equivalence. In our setting, this behavioural equivalence relies on a notion of static equivalence that captures indistinguishable sequences of messages.

**Definition 1.** *Two frames $\phi$ and $\psi$ are in* static equivalence, *$\phi \sim_s \psi$, when $\mathrm{dom}(\phi) = \mathrm{dom}(\psi)$, and $M\phi =_{\mathsf{E}} N\phi$ iff $M\psi =_{\mathsf{E}} N\psi$ for any recipes $M, N \in \mathcal{T}(\mathrm{dom}(\phi))$.*

Now, we lift this notion of equivalence from sequence of messages to configuration.

**Definition 2.** *Let $\mathcal{K}_P = (\mathcal{P}; \phi)$ and $\mathcal{K}_Q = (\mathcal{Q}; \psi)$ be two alive configurations with $\mathrm{dom}(\phi) = \mathrm{dom}(\psi)$. We write $\mathcal{K}_P \sqsubseteq_t \mathcal{K}_Q$ if for every execution $\mathcal{K}_P \xmapsto{\mathrm{tr}_1} (\mathcal{P}'; \phi')$, there exists $\mathrm{tr}_2$ and $(\mathcal{Q}'; \psi')$ such that $\mathcal{K}_Q \xmapsto{\mathrm{tr}_2} (\mathcal{Q}'; \psi')$, $\mathsf{obs}(\mathrm{tr}_1) = \mathsf{obs}(\mathrm{tr}_2)$ and $\phi' \sim_s \psi'$. Then, $\mathcal{K}_P \approx_t \mathcal{K}_Q$, if $\mathcal{K}_P \sqsubseteq_t \mathcal{K}_Q$ and $\mathcal{K}_Q \sqsubseteq_t \mathcal{K}_P$.*

We provide in Appendix A a description of the BAC protocol (e-passport) [5] in our model, together with the formalisation in terms of trace inclusion of the unlinkability property that it is supposed to achieve. We also recall there the action-determinism assumption required in previous works [8,9], and argue why it makes it impossible to adequately model privacy properties such as unlinkability. Similarly, the diff-equivalence used in ProVerif and Tamarin systematically leads to false attacks on unlinkability [27]. For such properties, one thus has to resort to verifying trace equivalence in the bounded setting but, the lack of POR techniques supporting non-action-deterministic processes is a major problem, since equivalence verification tools perform very poorly when the state explosion problem is left untamed.

## 3 Persistent and sleep sets in a nutshell

We review the key concepts of persistent and sleep sets, based on [26] but slightly reformulated. These general concepts apply to an action-deterministic LTS. We thus assume, in this section, a set of states $Q$, a set of actions $T$, and a partial transition function $\delta : Q \times T \to Q$. We write $s \xrightarrow{\alpha} s'$ when $s' = \delta(s, \alpha)$. We say that $\alpha$ is *enabled* in state $s$ if there exists an $s'$ such that $s' = \delta(s, \alpha)$. The set of enabled actions in $s$ is written $E(s)$. A state $s$ is *final* when $E(s) = \emptyset$.

**Definition 3.** *Independence is the greatest relation $\leftrightarrow \subseteq T \times Q \times T$ that is symmetric, irreflexive and such that, for all $(\alpha, s, \beta) \in \leftrightarrow$, also written $\alpha \leftrightarrow_s \beta$, we have:*
- *if $s \xrightarrow{\alpha} s'$ then $\beta \in E(s)$ iff $\beta \in E(s')$;*
- *if $s \xrightarrow{\alpha} s_1$ and $s \xrightarrow{\beta} s_2$, then $s_1 \xrightarrow{\beta} s'$ and $s_2 \xrightarrow{\alpha} s'$ for some $s'$.*

**Persistent sets.** A set $T \subseteq E(s)$ is *persistent in $s$* if, for all non-empty sequences of actions $s = s_0 \xrightarrow{\alpha_0} s_1 \ldots s_n \xrightarrow{\alpha_n} s_{n+1}$ such that $\alpha_i \notin T$ for all $0 \leq i \leq n$, we have that $\alpha_n \leftrightarrow_{s_n} \alpha$ for all $\alpha \in T$. We may note that $E(s)$ is persistent in $s$. In practice, persistent sets may be computed from *stubborn* sets (see Appendix B).

In the following, we assume a function $\mathsf{p_{set}} : Q \times T^* \to 2^T$ which associates to any state $s \in Q$ and any sequence $w$ such that $s \xrightarrow{w} s'$ with $E(s') \neq \emptyset$, a non-empty set of actions which is persistent in $s'$. A trace $s_0 \xrightarrow{\alpha_0} s_1 \ldots \xrightarrow{\alpha_n} s_{n+1}$ is *persistent*, written $s_0 \xrightarrow{\alpha_0 \ldots \alpha_n}_{\mathsf{p_{set}}} s_{n+1}$, if $\alpha_i \in \mathsf{p_{set}}(s_0, \alpha_0 \ldots \alpha_{i-1})$ for all $0 \leq i \leq n$.

**Proposition 1.** *Let $s'$ be a final state that is reachable from $s$. We have that $s'$ is also reachable from $s$ through a trace that is persistent.*

**Sleep sets.** If a persistent set contains two independent actions, then the associated search has redundancies. This has lead to the introduction of sleep sets that are formulated below. The technique relies on an arbitrary ordering $<$ on actions. A sleep set execution is an execution $(s_0, \emptyset) = (s_0, z_0) \xrightarrow{\alpha_0} (s_1, z_1) \ldots \xrightarrow{\alpha_n} (s_{n+1}, z_{n+1})$ with states in $Q \times 2^T$ such that $s_0 \xrightarrow{\alpha_0 \ldots \alpha_n}_{\mathsf{pset}} s_{n+1}$, and for any $0 \leq i \leq n$ we have $\alpha_i \notin z_i$ and $z_{i+1} = \{\beta \in z_i \mid \alpha_i \leftrightarrow_{s_i} \beta\} \cup \{\beta \in \mathsf{pset}(s_0, \alpha_0 \ldots \alpha_{i-1}) \mid \beta < \alpha_i,\ \alpha_i \leftrightarrow_{s_i} \beta\}$.

**Proposition 2.** *If a final state is reachable from $s$ in the original LTS, then it is also reachable from $(s, \emptyset)$ through a sleep set execution.*

## 4 Concrete LTS for security protocols

In order to apply the POR techniques of Section 3, we need to reformulate trace equivalence as a reachability property of final states in some LTS.

Given a set of handles $W \subseteq \mathcal{W}$, we define $\mathsf{Conf}(W)$ as the set of alive and *quiescent* configurations with a frame of domain $W$. An alive configuration $(\mathcal{P}; \phi)$ is quiescent if any $P \in \mathcal{P}$ is of the form $\mathsf{in}(c, x).P'$ or $\mathsf{out}(c, t).P'$ (in other words, no $\tau$ action can be triggered from it). We define $\mathsf{Conf}_\bot(W) = \{(\bot_j; \phi) \mid \mathsf{dom}(\phi) \subseteq W \text{ and } j \in \mathbb{N}\}$.

We define our *trace equivalence LTS* as follows:
- States are of the form $\langle|\mathbb{A} \approx \mathbb{B}|\rangle$ where $\mathbb{A}, \mathbb{B} \subseteq \mathsf{Conf}(W) \cup \mathsf{Conf}_\bot(W)$ for some $W \subseteq \mathcal{W}$, and at least one configuration in $\mathbb{A} \cup \mathbb{B}$ is alive. The *domain* $\mathsf{dom}(s)$ of such a state is $W$, and its *age* is $\mathsf{age}(s) = \max(\{0\} \cup \{j+1 \mid (\bot_j, \phi) \in \mathbb{A} \cup \mathbb{B}\})$.
- Actions are of the form $\mathsf{out}(c, \mathsf{w}_{c,i})$ or $\mathsf{in}(c, M)$ with $c \in \mathcal{Ch}, i \in \mathbb{N}, M \in \mathcal{T}(\mathcal{W})$.
- The transition relation is given by
$$s = \langle|\mathbb{A} \approx \mathbb{B}|\rangle \xrightarrow{\alpha} \langle|\mathbb{A}_g \uplus \mathbb{A}_n \uplus \mathbb{A}_a \approx \mathbb{B}_g \uplus \mathbb{B}_n \uplus \mathbb{B}_a|\rangle$$
where $\mathbb{A}_g, \mathbb{A}_a, \mathbb{A}_n$ are given below (and $\mathbb{B}_g, \mathbb{B}_n$, and $\mathbb{B}_a$ are defined similarly):
  - $\mathbb{A}_g = \mathbb{A} \cap \mathsf{Conf}_\bot(\mathsf{dom}(s))$,
  - $\mathbb{A}_a = \{A' \mid \exists A \in \mathbb{A} \text{ such that } A \xmapsto{\alpha} A'' \xmapsto{\tau^*} A' \not\xmapsto{\tau}\}$, and
  - $\mathbb{A}_n = \{(\bot_{\mathsf{age}(s)}; \phi) \mid (\mathcal{P}; \phi) \in \mathbb{A}, (\mathcal{P}; \phi) \text{ is alive}, (\mathcal{P}; \phi) \not\xmapsto{\alpha}\}$.

The transitions gather all alternatives that can perform the same output (resp. input) action. Therefore, even if our protocol allows several alternatives for a given observable action, our resulting trace equivalence LTS is action-deterministic. Configurations that cannot execute such an action become ghosts. A ghost configuration $(\bot_i; \phi)$ is a configuration that cannot evolve anymore, and the index $i$ is a crucial piece of information one needs to keep to know what other frames were present at that time (see Example 3).

Given a set of configurations $\mathbb{A}$, we define $\mathbb{A}^{\geq i}$ as the set of all configurations of $\mathbb{A}$ that are still alive at age $i$. More formally, we have that:
$$\mathbb{A}^{\geq i} = \{(\mathcal{P}, \phi) \in \mathbb{A} \mid (\mathcal{P}, \phi) \text{ is alive or } \mathcal{P} = \bot_j \text{ with } j \geq i\}$$
We write $\phi \sqsubseteq_s \psi$ when $\mathsf{dom}(\phi) \subseteq \mathsf{dom}(\psi)$ and both frames are in static equivalence on their common domain, *i.e.*, $\phi \sim_s \psi|_{\mathsf{dom}(\phi)}$. We lift $\sqsubseteq_s$ to a set of frames (and thus configurations) as follows: $\phi \sqsubseteq_s \Psi$ when there exists $\psi \in \Psi$ such that $\phi \sqsubseteq_s \psi$.

**Definition 4.** *A state $s = \langle|\mathbb{A} \approx \mathbb{B}|\rangle$ is* left-bad *when there exists $(\mathcal{P}; \phi) \in \mathbb{A}$ such that:*
- *either $(\mathcal{P}; \phi)$ is a ghost, i.e., $\mathcal{P} = \bot_j$ for some $j$, and $\phi \not\sqsubseteq_s \mathbb{B}^{\geq j}$;*

– or $(\mathcal{P}; \phi)$ is alive and $\phi \not\sqsubseteq_s (\mathbb{B} \cap \mathsf{Conf}(\mathrm{dom}(s)))$.

*The notion of being* right-bad *is defined similarly, and we say that a state $s$ is* bad *when it is right-bad or left-bad.*

We will see that trace non-equivalence implies the existence of a bad state. Thanks to ghosts, this will directly imply the existence of a final bad state. Fundamentally, ghosts are there to avoid that partial-order reduction makes us miss a bad state by not exploring certain transitions. Of course, practical verification algorithms will never perform explorations past a state that corresponds to a non-equivalence witness. Note, however, that detecting such states is only possible thanks to complex constraint solving, which we cannot afford in our symbolic POR algorithms. Hence, one important aspect in our design of ghosts is that they lift well to the "unsolved" symbolic setting.

*Example 3.* Ghosts are crucial to make sure that progressing in the LTS never kills a witness of non-equivalence. For instance, consider the two processes:
$P_u = \mathsf{out}(c, u) + (\mathsf{out}(c, n).\mathsf{out}(d, n'))$ where $u \in \{\mathsf{a}, \mathsf{b}\}$ are two public constants.
Consider $s_0 = \langle |(P_\mathsf{a}; \emptyset) \approx (P_\mathsf{b}; \emptyset)| \rangle$ and $s_0 \xrightarrow{\mathsf{out}(c, \mathsf{w}_{c,0})} s_1 \xrightarrow{\mathsf{out}(d, \mathsf{w}_{d,0})} s_2$ where:
 – $s_1 = \langle |\{(0; \{\mathsf{w}_{c,0} \triangleright \mathsf{a}\}), A\} \approx \{(0; \{\mathsf{w}_{c,0} \triangleright \mathsf{b}\}), A\}| \rangle$
 – $s_2 = \langle |\{(\bot_0; \{\mathsf{w}_{c,0} \triangleright \mathsf{a}\}), A'\} \approx \{(\bot_0; \{\mathsf{w}_{c,0} \triangleright \mathsf{b}\}), A'\}| \rangle$
 – $A = (\mathsf{out}(d, n'); \{\mathsf{w}_{c,0} \triangleright n\})$, and $A' = (0; \{\mathsf{w}_{c,0} \triangleright n, \mathsf{w}_{d,0} \triangleright n'\})$.
Note that $s_1$ is (left-)bad because $\{\mathsf{w}_c^0 \triangleright \mathsf{a}\} \not\sim_s \{\mathsf{w}_c^0 \triangleright \mathsf{b}\}$ and $s_2$ is (left-)bad because the ghost configurations are not statically equivalent either. However, without the ghost configurations, $s_2$ would not be bad (neither left nor right).

The following result reduces trace equivalence to reachability of a final bad state in the trace equivalence LTS, on which the POR techniques of Section 3 can be applied.

**Proposition 3.** *Let $A_0$ and $B_0$ be two alive configurations of same domain, and $s_0 = \langle |\mathbb{A}_0 \approx \mathbb{B}_0| \rangle$ where $\mathbb{A}_0 = \{A \mid A_0 \xmapsto{\tau^*} A \not\xmapsto{\tau}\}$, and $\mathbb{B}_0 = \{B \mid B_0 \xmapsto{\tau^*} B \not\xmapsto{\tau}\}$. The following conditions are equivalent:*

1. *$A_0$ is trace included in $B_0$, i.e., $A_0 \sqsubseteq_t B_0$;*
2. *no left-bad state is reachable from $s_0$ in the trace-equivalence LTS;*
3. *no left-bad, final state is reachable from $s_0$ in the trace-equivalence LTS.*

## 5 POR in symbolic semantics

The POR techniques of Section 3 apply to the LTS of Section 4, but this is not directly usable in practice because our trace equivalence LTS is infinitely branching. Symbolic execution is typically used to circumvent such problems, both in traditional software verification [11] and security protocol analysis [17,18]. In this section, we define a symbolic abstraction of our trace equivalence LTS, and we show how it can be used to effectively apply the persistent and sleep set techniques.

## 5.1 Symbolic equivalence LTS

As is common in symbolic semantics for security protocols [17,18], we rely on *second-order variables*, which will be instantiated by recipes, and *first-order variables*, which will be instantiated by messages. First-order variables are distinct from standard variables occurring in processes to represent input messages. More precisely, when an input is executed symbolically, the associated variable will be substituted by a first-order variable. As a result, standard variables will only occur bound in symbolic processes, while first-order variables will only occur free. Conversely, only first-order variables will be allowed to occur free in processes, frames, and states.

Second-order and first-order variables will respectively be of the form $\mathsf{X}^{c,i}$ and $\mathsf{x}^{c,i}_\phi$ where $c \in \mathcal{C}h$, $i \in \mathbb{N}$, and $\phi$ is a symbolic frame, *i.e.*, a frame whose terms may contain first-order variables. Intuitively, $\mathsf{X}^{c,i}$ stands for the recipe used for the $i^{\text{th}}$ input on channel $c$, and $\mathsf{x}^{c,i}_\phi$ will be instantiated by the message resulting from that recipe in the context of the frame $\phi$. The use of variables with explicit $c, i$ parameters avoids us to deal with freshness or $\alpha$-renaming issues when implementing the symbolic analysis. We denote $vars^1(t)$ (resp. $vars^2(t)$) the first-order (resp. second-order) variable occurring in $t$. Finally, $vars(R)$ is the set of handles that occur in a recipe $R$. We say that a symbolic frame $\phi$ is *well-founded* if, whenever $\phi(\mathsf{w}_{c,i}) = t$ and $\mathsf{x}^{d,j}_\psi \in vars^1(t)$, it holds that $\psi \subsetneq \phi$. This well-foundedness condition will obviously be preserved in symbolic executions: if $t$ is the $i^{\text{th}}$ output on channel $c$, it may only depend on inputs received before that output, *i.e.*, at a time where the frame $\psi$ does not contain $\mathsf{w}_{c,i}$. From now on, we impose that all frames are well-founded, which allows us to define the first-order substitution associated to a second-order substitution.

**Definition 5** ($\lambda_\theta$). *Let $\theta$ be a substitution mapping second-order variables to recipes. Its associated first-order substitution $\lambda_\theta$ is the unique substitution of (infinite) domain $\{\mathsf{x}^{c,i}_\phi \mid vars(\mathsf{X}^{c,i}\theta) \subseteq \mathrm{dom}(\phi)\}$ such that $\lambda_\theta(\mathsf{x}^{c,i}_\phi) = (\mathsf{X}^{c,i}\theta)(\phi\lambda_\theta)$, which can be defined by induction on the size of frame domains.*

We now define symbolic actions and states, and their concretisations. We take *symbolic actions* of the form $\mathtt{out}(c, \mathsf{w}_{c,i})$ and $\mathtt{in}(c, \mathsf{X}^{c,i}, W)$, where $c \in \mathcal{C}h$, $i \in \mathbb{N}$ and $W \subseteq \mathcal{W}$. Given a substitution $\theta$ mapping second-order variables to recipes, we define the $\theta$-concretisations of symbolic actions as follows: $\mathtt{out}(c, \mathsf{w}_{c,i})\theta = \mathtt{out}(c, \mathsf{w}_{c,i})$, and $\mathtt{in}(c, \mathsf{X}^{c,i}, W)\theta = \mathtt{in}(c, R)$ when $\mathsf{X}^{c,i}\theta = R \in \mathcal{T}(W)$.

We will use constraints which are conjunctions of equations and disequations over (open) terms. The empty constraint is written $\top$, and conjunction is written $\wedge$ and considered modulo associativity-commutativity.

A *symbolic state* $S = \langle \mathbb{A} \approx \mathbb{B} \rangle^I_\mathcal{C}$ is formed from a mapping $I : \mathcal{C}h \to \mathbb{N}$ providing input numbers, a constraint $\mathcal{C}$, and two sets $\mathbb{A}$ and $\mathbb{B}$ of *symbolic configurations*, *i.e.*, configurations that may contain first-order variables. We further require that:
- there is at least one alive configurations in $\mathbb{A} \cup \mathbb{B}$;
- all alive configurations of $\mathbb{A} \cup \mathbb{B}$ share the same frame domain, noted $\mathrm{dom}(S)$;
- any ghost configuration in $\mathbb{A} \cup \mathbb{B}$ should have a domain $W \subseteq \mathrm{dom}(S)$;
- processes in configurations do not contain null processes, and do not feature top-level conditionals, parallel and choice operators.

With this in place, we define the solutions of $S = \langle \mathbb{A} \approx \mathbb{B} \rangle_\mathcal{C}^I$ as the set $\mathsf{Sol}(S)$ containing all the substitutions $\theta$ such that:
- $\mathrm{dom}(\theta) = \{\mathsf{X}^{c,i} \mid i < I(c)\}$ and $vars^1(S) \subseteq \mathrm{dom}(\lambda_\theta)$;
- for any $u = v$ (resp. $u \neq v$) in $\mathcal{C}$, $u\lambda_\theta =_\mathsf{E} v\lambda_\theta$ (resp. $u\lambda_\theta \neq_\mathsf{E} v\lambda_\theta$).

Given a state $S$, and $\theta \in \mathsf{Sol}(S)$, we define its $\theta$-concretisation $S\theta$ as $\langle |\mathbb{A}\lambda_\theta \approx \mathbb{B}\lambda_\theta| \rangle$.

*Remark 1.* Beyond the differences in formalism, our notion of solution is quite close to ones found, *e.g.*, in [17,18], with one difference: when no $\mathsf{x}_\phi^{c,i}$ variable occurs in $S$, $\theta(\mathsf{X}^{c,i})$ is completely unconstrained. This means that when an input variable is unused in the input's continuations, our solutions are incorrect wrt. the corresponding recipe. We do not need to worry about this mismatch, though, because we only need a symbolic semantics that covers all concrete executions; it does not need to be sound. In fact, our analysis will never rely on the existence of a solution for a given symbolic state. It will never check that a term is deducible, and will almost ignore (dis)equality constraints, only checking for immediate contradictions among them.

We can now define symbolic transitions, and establish their completeness.

**Definition 6.** *Consider a symbolic state $S = \langle \mathbb{A} \approx \mathbb{B} \rangle_\mathcal{C}^I$ and a symbolic action $A$, the possible transitions $S \xrightarrow{A} S'$ are defined by mimicking concrete transitions as follows:*
- *We first execute the action $A$, gathering all possible resulting configurations into a pre-state $S_A = \langle \mathbb{A}' \approx \mathbb{B}' \rangle_{\mathcal{C}'}^{I'}$. To be possible, such a transition has to be of the form $A = \mathsf{in}(c, \mathsf{X}^{c,i}, W)$ with $i = I(c)$, or $A = \mathsf{out}(c, \mathsf{w}_{c,i})$ with $i = \#_c(\mathrm{dom}(S))$. The resulting pre-state $S_A$ is not a valid state because it may contain e.g., top-level conditionals, choice operators. This pre-state also includes ghosts $(\bot_n; \phi)$ of the configurations $(\mathcal{P}; \phi)$ of $S$ that could not perform $A$, where $n = \mathsf{age}(S)$ as defined in the concrete semantics. We define $I'$ to coincide with $I$ on all channels, except on $c$ where $I'(c) = I(c) + 1$ when $A$ is an input on $c$. When executing $A = \mathsf{in}(c, \mathsf{X}^{c,i}, W)$ in a configuration $(\mathcal{P}; \phi)$ of $S$ that can perform an input on $c$, we use the term $\mathsf{x}_{\phi|_W}^{c,i}$ to substitute for the input variable.*
- *Then we declare $S \xrightarrow{A} S'$ if $S'$ is a state that can be obtained from $S_A$ by repeatedly performing the following operations, until none applies:*
  - *If a configuration features a top-level conditional, the conditional is replaced by one of its branches, and the constraints are enriched accordingly.*
  - *If a configuration features a top-level choice operator, it is replaced by the two configurations where the choices are made.*
  
  *We also require that $S'$ does not have an immediately contradicting constraint, i.e., a constraint containing an equation and its negation.*

A perhaps surprising consequence of our definition is that, if $\mathsf{in}(c, \mathsf{X}^{c,i}, W)$ is executable in $S$, then any $\mathsf{in}(c, \mathsf{X}^{c,i}, W')$ is also executable. Allowing smaller domains is important for checking independencies. We also allow larger domains, possibly even larger than $\mathrm{dom}(S)$, mainly because it simplifies the theory, at no cost in practice.

*Example 4.* Consider arbitrary terms $t$, $u$, and $v \neq v'$, and the symbolic state $S = \langle (\mathcal{P}; \psi) \approx (\mathcal{P}; \psi') \rangle_\top^I$ where $\mathcal{P} = \mathsf{in}(c, x).\mathsf{if}\ x = t\ \mathsf{then}\ \mathsf{out}(c, \mathsf{ok})\ \mathsf{else}\ 0$,

$$\phi = \{\mathsf{w}_{c,0} \mapsto u\}, \quad \psi = \phi \uplus \{\mathsf{w}_{d,0} \mapsto v\} \text{ and } \psi' = \phi \uplus \{\mathsf{w}_{d,0} \mapsto v'\}.$$

We illustrate how the choice of $W$ affects which transitions are possible from state $S$ with action $A = \mathtt{in}(c, \mathsf{X}^{c,i}, W)$, where $i = I(c)$ is the only value that allows this action to execute, and $I'$ coincides with $I$ except on $c$ for which $I'(c) = I(c) + 1$. If $W = \{\mathsf{w}_{c,0}\}$, then there are two possible transitions:

$$S \xrightarrow{A} \langle(\mathtt{out}(c,\mathsf{ok}); \psi) \approx (\mathtt{out}(c,\mathsf{ok}); \psi')\rangle^{I'}_{\mathsf{x}^{c,i}_\phi = t} \quad S \xrightarrow{A} \langle(0; \psi) \approx (0; \psi')\rangle^{I'}_{\mathsf{x}^{c,i}_\phi \neq t}$$

If $W = \{\mathsf{w}_{c,0}, \mathsf{w}_{d,0}\}$, four transitions are possible, notably including

$$S \xrightarrow{A} \langle(\mathtt{out}(c,\mathsf{ok}); \psi) \approx (0; \psi')\rangle^{I'}_{\mathsf{x}^{c,i}_\psi = t,\, \mathsf{x}^{c,i}_{\psi'} \neq t}.$$

Indeed, we are considering here an input whose recipe may exploit the different frames of our two configurations. It is a priori possible that the resulting message passes the test $x = t$ only in one configuration.

*Remark 2.* Symbolic executions preserve various properties that we do not exploit, but may be useful to state for intuitions. In a configuration $(\mathcal{P}; \phi)$ of a state $\langle \mathbb{A} \approx \mathbb{B} \rangle^I_\mathcal{C}$, the only first-order variables that appear are of the form $\mathsf{x}^{c,i}_\psi$ with $\psi \subseteq \phi$ and $i < I(c)$.

We note $\theta' \sqsupseteq \theta$ when $\theta'$ is an extension of $\theta$, *i.e.*, $\theta'|_{\mathrm{dom}(\theta)} = \theta$.

**Proposition 4.** *Let $S = \langle \mathbb{A} \approx \mathbb{B} \rangle^I_\mathcal{C}$ be a symbolic state, $\theta \in \mathsf{Sol}(S)$. Let $s'$ and $\alpha$ be such that $S\theta \xrightarrow{\alpha} s'$. There exists $S'$, $A$ and $\theta' \sqsupseteq \theta$ such that $S \xrightarrow{A} S'$, $\theta' \in \mathsf{Sol}(S')$, $\alpha = A\theta'$, and $s' = S'\theta'$. Moreover, if $\alpha$ is of the form $\mathtt{in}(c, R)$, the proposition holds with $A = \mathtt{in}(c, \mathsf{X}^{c,I(c)}, W)$ for any $W$ such that $vars(R) \subseteq W$.*

### 5.2 Independence relations

We first define the *enabled symbolic independence* relation, and show that it is a sound abstraction of independence for enabled actions. For that, we assume here a notion of *incompatible* constraints. It can be anything as long as two constraints $\mathcal{C}$ and $\mathcal{C}'$ are only declared incompatible when $\mathcal{C} \wedge \mathcal{C}'$ is unsatisfiable. In practice, we only check for immediate contradictions, *i.e.*, the presence of an equation and its negation. This allows us to easily check $\Leftrightarrow^{ee}$ in the implementation.

**Definition 7.** *Given a symbolic state $S$, and two symbolic actions $A$ and $B$ executable in $S$, we write $A \Leftrightarrow^{ee}_S B$ when:*
- *$A$ and $B$ are neither two inputs nor two outputs on the same channel;*
- *for any $S \xrightarrow{A} S_A$, $S \xrightarrow{B} S_B$, we have that $S_A \xrightarrow{B} S_{AB}$ and $S_B \xrightarrow{A} S_{BA}$ for some symbolic states $S_{AB}$ and $S_{BA}$;*
- *for any $S \xrightarrow{A} S_A \xrightarrow{B} S_{AB}$, and $S \xrightarrow{B} S_B \xrightarrow{A} S_{BA}$, we have that $S_{AB}$ and $S_{BA}$ have incompatible constraints, or $S_{AB} = S_{BA}$.*

We now turn to defining a sound abstraction of independence between a concretely disabled and enabled action. Intuitively, $A \Leftrightarrow^{de}_S B$ will guarantee that executing concretisations of $B$ cannot enable new concretisations of $A$.

**Definition 8.** *Given a symbolic state $S$, as well as two symbolic actions $A$ and $B$, we write $A \Leftrightarrow^{de}_S B$ when $B$ is executable in $S$, and*

- *either $A$ is not executable in $S'$ for any $S'$ such that $S \overset{B}{\twoheadrightarrow} S'$;*
- *or $A$ is executable in $S$ but $A/B$ are not of the form $\texttt{in}(c, \mathsf{X}^{c,i}, W)/\texttt{out}(d, \mathsf{w}_{d,j})$ with $\mathsf{w}_{d,j} \in W$.*

**Proposition 5.** *Let $S$ be a symbolic state and $A$ and $B$ be two symbolic actions. Let $\theta \in \mathsf{Sol}(S)$, $s = S\theta$ and $\alpha$ (resp. $\beta$) be a concretisation of $A$ (resp. $B$).*
- *If $A \Leftrightarrow_S^{ee} B$, and $\alpha, \beta \in E(s)$, then $\alpha \leftrightarrow_s \beta$.*
- *If $A \Leftrightarrow_S^{de} B$, $\alpha \notin E(s)$ and $\beta \in E(s)$, then $\alpha \leftrightarrow_s \beta$.*

*Example 5.* Let $\mathcal{P} = \texttt{in}(c,x).\texttt{out}(c,x) \mid \texttt{out}(d,t)$, and $S = \langle (\mathcal{P}; \emptyset) \approx (\mathcal{P}; \emptyset) \rangle_\top^{I_0}$ with $I_0(c) = 0$ for any $c \in \mathcal{C}h$. We have $\texttt{in}(c, \mathsf{X}^{c,0}, \emptyset) \Leftrightarrow_S^{ee} \texttt{out}(d, \mathsf{w}_{d,0})$: inputs and outputs commute, for inputs whose recipes rely on the currently available (empty) domain. We have $\texttt{in}(c, \mathsf{X}^{c,0}, \emptyset) \Leftrightarrow_S^{de} \texttt{out}(d, \mathsf{w}_{d,0})$ (the output does not enable new concretisations for the input) but *not* $\texttt{in}(c, \mathsf{X}^{c,0}, \{\mathsf{w}_{d,0}\}) \Leftrightarrow_S^{de} \texttt{out}(d, \mathsf{w}_{d,0})$ (the input is feasible, but performing it after the output would enable new concretisations).

### 5.3 Persistent set computation

Having defined over-approximations of transitions and dependencies, we now describe how to compute, for a state $S$, a set of actions $T^+(S)$ that yields a persistent set for any concretisation of $S$. More precisely, we shall compute stubborn sets (cf. Appendix B.2).

Our symbolic LTS is still infinitely branching, due to the absence of constraints on inputs domains $W$. However, when exploring the LTS, it often suffices to consider inputs with a canonical domain, *i.e.*, the domain of the current state. We formalise this by defining the *enabled cover* of a symbolic state $S$: $EC(S)$ is the set of all actions that are executable in $S$, with the constraint that inputs are of the form $\texttt{in}(c, \mathsf{X}^{c,i}, \mathrm{dom}(S))$. Proposition 4 already ensures that any concrete action in $E(S\theta)$ can be mapped to a symbolic action in $EC(S)$.

**Definition 9.** *Let $S$ be a symbolic state, $A$ and $B$ be two symbolic actions such that $B$ is executable in $S$. We say that $A \Leftrightarrow_S B$ when (i) $A \Leftrightarrow_S^{de} B$ and, (ii) if $A$ is executable in $S$ then $A \Leftrightarrow_S^{ee} B$.*

Given a symbolic state $S$, we say that a set of actions $X$ is a *symbolic stubborn set for $S$* when $X \cap EC(S) \neq \emptyset$ and, for any $A \in X$ and any execution
$$S = S_1 \overset{B_1}{\twoheadrightarrow} S_2 \ldots S_n \overset{B_n}{\twoheadrightarrow} S_{n+1} \text{ with } B_i \in EC(S_i) \text{ for all } 1 \leq i \leq n$$
such that $A \not\Leftrightarrow_{S_n} B_n$, there exists $1 \leq i \leq n$ such that $B_i \in X$.

We assume a computable function which associates to any symbolic state $S$ such that $EC(S) \neq \emptyset$ a set $T^+(S)$ that is a symbolic stubborn set for $S$. Computing $T^+(S)$ is typically achieved as a least fixed point computation, initialising the set with an arbitrary action in $EC(S)$, exploring executions that avoid the current set and adding actions $B_n$ when they are dependent with an action already in the set. In this process all transitions in the enabled cover of $S$ and its successors are considered (unless they are in the current set) without caring for the existence of a solution for the visited states. The computation is carried out with each possible action of $EC(S)$ as its initial set, and a result of minimal cardinality is kept. In the worst case, it will be $EC(S)$ itself.

If done in a depth-first fashion, the computation is (a symbolic approximation of) Godefroid's stubborn set computation through first conflict relations [26]. It is however much more efficient to perform the explorations in breadth, since the addition of an action along an exploration can potentially prevent the continuation of another exploration. In any case, the details of how $T^+(S)$ is computed do not matter for correctness.

*Example 6.* Consider the process $P = \text{in}(c, x).Q \mid \text{in}(d, x).\text{out}(d, t).Q'$ where $Q$, $Q'$ and $t$ are arbitrary. Consider computing $T^+(S)$ for $S = \langle (P; \emptyset) \approx (P; \emptyset) \rangle_{\top}^{I_0}$, initializing the set with $A_0 = \text{in}(c, \mathsf{X}^{c,0}, \emptyset)$.

Since $A_0 \Leftrightarrow_S \text{in}(d, \mathsf{X}^{d,0}, \emptyset)$ we have to explore successors of $S$ by the input on $d$. There is only one, call it $S'$. We have $A_0 \Leftrightarrow_{S'} \text{out}(d, \mathsf{w}_{d,0})$, so again we consider the successor $S''$ by the output action. We have $A_1 = \text{in}(c, \mathsf{X}^{c,0}, \{\mathsf{w}_{d,0}\}) \in EC(S'')$ with $A_1 \not\Leftrightarrow_{S''} A_0$, hence we add $A_1$ to our set. We repeat the process from $S$, exploring the transition along $\text{in}(d, \mathsf{X}^{d,0}, \emptyset) \Leftrightarrow_S A_1$. But, we then have $\text{out}(d, \mathsf{w}_{d,0}) \not\Leftrightarrow_{S'} A_1$, and more precisely $A_1 \not\Leftrightarrow_{S'}^{de} \text{out}(d, \mathsf{w}_{d,0})$. Hence we add $A_2 = \text{out}(d, \mathsf{w}_{d,0})$ to our set. Because $A_2 \not\Leftrightarrow_S^{de} \text{in}(d, \mathsf{X}^{d,0}, \emptyset) = A_3$, we will also add that action in the next iteration. We thus obtain $T^+(S) = \{A_0, A_1, A_2, A_3\}$, satisfying our specification of $T^+$. This symbolic stubborn set yields the symbolic persistent set $T^+(S) \cap EC(S) = \{\text{in}(c, \mathsf{X}^{c,0}, \emptyset), \text{in}(d, \mathsf{X}^{d,0}, \emptyset)\}$; in that case, no reduction is possible. However, starting with process $P \mid \text{out}(e, t').P'$ and initialising the set with $A_4 = \text{out}(e, \mathsf{w}_{e,0})$ will often lead to a very good reduction, *i.e.*, a singleton. This is not always the case, however, due to non-determinism and ghosts (*i.e.*, when an output on $e$ occurs in $Q$ or $Q'$, or when an action in another configuration creates a ghost).

**Proposition 6.** *Let $S$ be a symbolic state such that $EC(S) \neq \emptyset$, and consider $T = \{A\theta \mid A \in T^+(S)\}$. For any $\theta' \in \mathsf{Sol}(S)$, the set $T \cap E(S\theta')$ is persistent in $S\theta'$.*

Having computed symbolic persistent sets, we now define a persistent set assignment $\mathsf{p_{set}}$ for the concrete LTS. By completeness, we know that, for any concrete execution $s_0 \xrightarrow{\alpha_0} s_1 \ldots \xrightarrow{\alpha_{n-1}} s_n$ there exists $S_0 \xrightarrow{A_0} S_1 \ldots \xrightarrow{A_{n-1}} S_n$ and $\theta_0 \sqsubseteq \theta_1 \ldots \sqsubseteq \theta_n$ with $\theta_0$ the empty substitution, $\theta_i \in \mathsf{Sol}(S_i)$ and $S_i\theta_i = s_i$ for all $i \in [0; n]$, and $A_i\theta_{i+1} = \alpha_i$ for all $i \in [0; n-1]$. We assume a choice function $\mathsf{abs}$ which, to each such concrete execution associates a symbolic abstraction: $\mathsf{abs}(s_0, \alpha_0 \ldots \alpha_{n-1}) = (S_0, S_1, \ldots, S_n)$. We can assume that the choice is compatible with prefixing:

$\mathsf{abs}(s_0, \alpha_0 \ldots \alpha_n) = (S_i)_{0 \leq i \leq n+1}$ implies $\mathsf{abs}(s_0, \alpha_0 \ldots \alpha_{n-1}) = (S_i)_{0 \leq i \leq n}$.

Building on this, we define

$$\mathsf{p_{set}}(s_0, \alpha_0 \ldots \alpha_{n-1}) = \{A\theta \mid A \in T^+(S_n)\} \cap E(S_n\theta_n)$$

where $\mathsf{abs}(s_0, \alpha_0 \ldots \alpha_{n-1}) = (S_0, \ldots, S_n)$, which, by Proposition 6, is a persistent set in $s_n$ (uniquely defined as the state reachable from $s_0$ after $\alpha_0 \ldots \alpha_{n-1}$). In other words, we obtain the persistent set for a concrete state from the symbolic persistent set of one of its symbolic abstractions, but we choose this abstraction depending on the concrete execution and not only its resulting state.

With this in place, Proposition 1 guarantees that for any execution from $s_0$ to a final state $s_f$, there exists a persistent execution (wrt. $\mathsf{p_{set}}$) from $s_0$ to $s_f$. Hence, the search for final bad states can be restricted to only explore concretisations of *symbolic* *persistent traces*, *i.e.*, symbolic executions where the only transitions considered for a state $S$ are those in $T^+(S) \cap EC(S)$. We will make this more formal after having applied sleep sets.

### 5.4 Symbolic sleep sets

We finally describe how we implement sleep sets symbolically. We shall define a symbolic LTS with sleep sets, whose states $(S, Z)$ compound a symbolic state $S$ and a set of symbolic actions $Z$. The sleep set technique relies on a strict ordering of actions, but the order is only relevant for comparing independent actions, which do not have the same skeleton (the skeleton of an action denotes its input/output nature and its channel). Thus, we assume a strict total order $<$ on action skeletons, and lift it to symbolic and concrete actions.

Then, a sleep set execution in our symbolic LTS is any execution
$$(S_0, \emptyset) = (S_0, Z_0) \stackrel{A_0}{\twoheadrightarrow} (S_1, Z_1) \ldots (S_n, Z_n) \stackrel{A_n}{\twoheadrightarrow} (S_{n+1}, Z_{n+1})$$
such that for $0 \leq i \leq n$, we have that $A_i \in T^+(S_i) \cap EC(S_i)$, $A_i \notin Z_i$, and $Z_{i+1} = \{B \in Z \mid B \Leftrightarrow^{ee}_{S_i} A_i\} \cup \{A' \in T^+(S_i) \cap EC(S_i) \mid A' < A_i,\ A' \Leftrightarrow^{ee}_{S_i} A_i\}$.

These symbolic sleep set executions are complete with respect to the sleep set technique applied to our concrete LTS with the $\mathsf{p}_{\mathsf{set}}$ function defined above.

**Proposition 7.** *Let* $(s_0, \emptyset) \stackrel{\alpha_0}{\longrightarrow} (s_1, z_1) \ldots \stackrel{\alpha_{n-1}}{\longrightarrow} (s_n, z_n)$ *be a sleep set execution in our initial LTS. Then, there is* $(S_0, \emptyset) \stackrel{A_0}{\twoheadrightarrow} (S_1, Z_1) \ldots \stackrel{A_{n-1}}{\twoheadrightarrow} (S_n, Z_n)$ *a sleep set execution in our symbolic LTS, and substitutions* $\emptyset = \theta_0 \sqsubseteq \theta_1 \ldots \sqsubseteq \theta_n$ *such that* $s_i = S_i \theta_i$, $\alpha_i = A_i \theta_{i+1}$ *for* $i \in [1; n-1]$, *and* $s_n = S_n \theta_n$.

*Example 7.* Let $S$ be the state from Example 6. Starting with $(S, \emptyset)$ we may perform two transitions in the sleep LTS: $A_0 = \mathtt{in}(c, \mathsf{X}^{c,0}, \emptyset)$ and $A_3 = \mathtt{in}(d, \mathsf{X}^{d,0}, \emptyset)$. Assuming that $A_0 > A_3$ and $S \stackrel{A_0}{\twoheadrightarrow} S_c$, we have $(S, \emptyset) \stackrel{A_0}{\twoheadrightarrow} (S, \{A_3\})$. Assuming now that $Q$ starts with another input on $c$, the persistent set for $S_c$ will contain inputs on $c$ and $d$. However, executing $A_3$ is not allowed in $(S, \{A_3\})$. Intuitively, while the persistent set technique only looks forward, the sleep set technique also takes into account the past, and indicates here that exploring $A_3$ is not useful after $A_0$, since it can equivalently be performed before it.

### 5.5 Collapsing conditionals

The above techniques allow us, in principle, to compute significantly reduced set of symbolic traces whose concretisations contain a witness of non-equivalence when such a witness exists. However, the algorithm for computing persistent sets is quite inefficient when applied on practical case studies: it relies on explorations of their symbolic LTS, which is highly branching and too large due to conditionals. This is a typical problem of symbolic execution, which manifests itself acutely in our setting, where the branching factor of a state is generally the product of those of its configurations. We circumvent this difficulty by observing that stubborn sets for a state (and its sleep set executions) can be computed by analysing a transformed state where conditionals are pushed down.

Our transformation can often completely eliminate conditionals in our case studies, and is key to obtaining acceptable performances.

To justify an elementary step of this transformation, we consider a symbolic state $S$ containing a conditional we would like to simplify: $S = S'[\text{if } u = v \text{ then } P \text{ else } Q]$ ($S'[\cdot]$ denotes a state with a hole). We require that $P$ and $Q$ are respectively of the form $\alpha.P'$ and $\beta.Q'$ where $\alpha$ and $\beta$ have the same skeleton. We make the observation that, independently of the execution and the evaluation of the test $u = v$, the same action will be released and, in case of outputs, the precise output term has little impact in the context of our symbolic analysis. Following this intuition, we would like to postpone the conditional by considering $S^c = S'[\gamma.\text{if } u = v \text{ then } P' \text{ else } Q']$, where $\gamma$ is either the input $\alpha = \beta$, or a well-chosen combination of the outputs $\alpha$ and $\beta$. The choice of $\gamma$ should ensure that the transformation cannot create action independencies that did not hold before the transformation.

Formally, we assume a fresh function symbol $\Delta$ of arity 4, and take $S^c = S'[T^c]$ where $T^c$ is defined as:
- $\text{in}(c, x).\text{if } u = v \text{ then } P' \text{ else } Q'$ when $(\alpha, \beta) = (\text{in}(c, x), \text{in}(c, x))$;
- $\text{out}(c, \Delta(t, t', u, v)).\text{if } u = v \text{ then } P' \text{ else } Q'$ when $(\alpha, \beta) = (\text{out}(c, t), \text{out}(c, t'))$.

**Proposition 8.** *For any execution $S = S_0 \overset{A_1}{\twoheadrightarrow} S_1 \ldots \overset{A_n}{\twoheadrightarrow} S_n$, there is an execution $S^c = T_0 \overset{A_1}{\twoheadrightarrow} T_1 \ldots \overset{A_n}{\twoheadrightarrow} T_n$, such that, for any $A$ and $i \in [1; n]$, $A \Leftrightarrow_{T_{i-1}} A_i$ (resp. $A_i \Leftrightarrow_{T_{i-1}} A$) implies $A \Leftrightarrow_{S_{i-1}} A_i$ (resp. $A_i \Leftrightarrow_{S_{i-1}} A$).*

*Hence, $T^+(S^c)$ is a symbolic stubborn set for $S$ and any sleep set execution from $S$ is also a sleep set execution from $S^c$.*

Repeatedly applying this result, and an obvious analogue justifying the elimination of conditionals whose branches are null, we can eliminate most conditionals from our protocols, and compute stubborn sets and sleep set executions efficiently.

## 6 Implementation and benchmarks

The results of the previous sections allow us to compute a set of symbolic actions, that can be used to restrict the search when looking for a witness of non-equivalence. By Proposition 3, $(P_1; \emptyset) \not\approx (P_2; \emptyset)$ iff a bad state can be reached from $s_0 = \langle |\mathbb{B}_1 \approx \mathbb{B}_2| \rangle$, where $\mathbb{B}_i = \{\mathcal{K}_i \mid (P_i; \emptyset) \overset{\tau^*}{\longmapsto} \mathcal{K}_i \overset{\mathcal{T}}{\not\longmapsto}\}$. By Proposition 2, this implies the existence of a sleep set execution in our trace equivalence LTS from $(s_0, \emptyset)$ to a bad state. By Proposition 7, this implies the existence of a concrete execution whose underlying symbolic trace $S_0 = \langle |\mathbb{B}_1 \approx \mathbb{B}_2 \rangle \overset{I_0}{\top} \overset{A_0}{\twoheadrightarrow} \ldots \overset{A_n}{\twoheadrightarrow} S_{n+1}$ is a sleep set execution in our symbolic LTS. Such symbolic traces can be computed.

### 6.1 Implementation

To concretely realise and evaluate our techniques, we have implemented our symbolic analysis as a standalone library called Porridge, and have interfaced it with Apte in the first place, and then with its successor DeepSec, once this tool has been made available [19]. These tools perform an exhaustive search for non-equivalence witnesses using symbolic execution. Conceptually, this search can be seen as a naive symbolic exploration, combined with an elaborate constraint solving procedure. The two aspects being orthogonal, we can straightforwardly obtain a correct optimisation by restricting the symbolic exploration according to the set of traces computed by Porridge.

**Porridge.** The library is open-source, written in OCaml, and available at [1]. The code implements exactly the techniques presented above, with only a few minor additions. It consists of ∼6k LoC. Performance-wise, we heavily make use of hashconsing and memoization, but not from multicore programming yet. The design of the library, with an independent POR functor, makes it easy to apply symbolic POR analyses to other LTS; we can already perform POR for trace inclusion, and expect to use this flexibility to adapt our POR techniques to slightly different protocol semantics.

**Integration in Apte and DeepSec.** As mentioned above, Apte and DeepSec are based on constraint solving procedures on top of which an exhaustive and naive symbolic executions exploration is performed. This exploration is naive in the sense that all interleavings are considered (except for the specific case of action-deterministic protocols already discussed in introduction). We have shown that restricting the exploration to symbolic sleep set traces still yields a decision procedure for trace equivalence. This restriction is easily implemented, as lightweight modifications (∼500 LoC) of Apte and DeepSec [1]. Note that the differences between the semantics presented in Section 2 and the ones used by those tools can easily be ignored by slightly restricting the class of protocols. Concretely, we exploit the class of protocols with non-blocking outputs as done in [9], which is not restrictive.

## 6.2 Experimental evaluation

We have carried out numerous benchmarks, focusing on DeepSec since it is both more general and more efficient than Apte, and measuring the improvements brought by Porridge in terms of computation time and number of explorations. The latter is also a good indicator of the effectiveness of the reduction achieved since it represents the number of times DeepSec explores an action and applies its costly constraint solving procedure.

**Case studies.** We verify some privacy properties on several real-life protocols of various sizes by modifying the number of sessions being analysed. We model unlinkability [27,5] of the BAC protocol [5] (used in e-passports, see Appendix A), of Private Authentication [3] and of Feldhofer [24]. We also model anonymity [27,5] for some of them. All our case studies are available at [1].

**Setup.** We run DeepSec and Porridge both compiled with OCaml 4.06.0 on a server running Ubuntu 16.04.5 (Linux 4.4.0) with 12∗2 Intel(R) Xeon(R) CPU E5-2650 v4 @ 2.20GHz and 256G of RAM. We run each test case on a single core with a time-out of 2 hours (real-time) and maximal memory consumption of 10G.

**Results.** We report in Table 1 the relative speed-up of computation time and the reduction of explorations brought by Porridge. We plot the same information for numerous sizes of Private Authentication in Figure 1. We observe that speed-ups are closely related to the reduction achieved on the number of explorations. As the size of protocols increases, Porridge quickly speeds up computations by more that 1 order of magnitude.

| Test | Size | Time (ratio) | Explorations (ratio) | Time (seconds) |
|---|---|---|---|---|
| BAC (UK) | 4 | 7.6 | 7.23 | 12.23 |
| Private Auth. (ANO) | 2 | 1.25 | 2.71 | 0.04 |
| Private Auth. (ANO) | 3 | 1.67 | 4.01 | 0.04 |
| Private Auth. (ANO) | 4 | 8.21 | 10.51 | 1.17 |
| Private Auth. (ANO) | 5 | 14.89 | 16.61 | 10.57 |
| Private Auth. (ANO) | 6 | 60.2 | 36.75 | 4864 |
| Private Auth. (UK) | 2 | 2.29 | 9.6 | 0.16 |
| Private Auth. (UK) | 3 | 14.06 | 29.77 | 79.57 |
| Private Auth. (UK) | 4 | 46.2 | 46.69 | 7171 |
| Feldhofer (ANO) | 2 | 1 | 4.72 | 0.03 |
| Feldhofer (ANO) | 3 | 4.63 | 7.08 | 0.37 |
| Feldhofer (ANO) | 4 | 22.47 | 16.3 | 544.93 |
| Feldhofer (UK) | 4 | 36.27 | 22.58 | 1510.09 |

**Table 1.** Relative speed-up and reduction of explorations with Porridge vs. without Porridge. In the last column, we show the computation time without Porridge. The size refers to the total number of processes in parallel. ANO (resp. UK) refers to Anonymity (resp. Unlinkability).

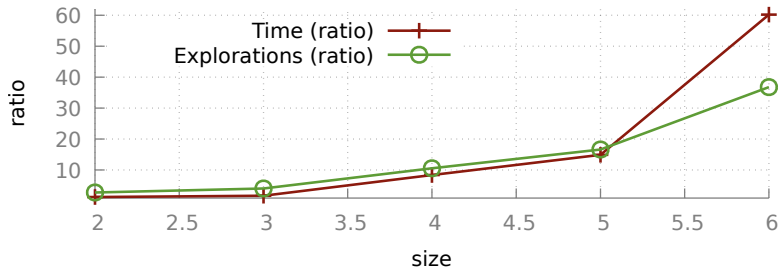

**Fig. 1.** Relative speed-up and reduction of explorations with Porridge vs. without Porridge on Private Authentication (ANO) of different sizes.

With each size increase, we add one role out of two, explaining the more significant increases for even values.

## 7 Conclusion

We have presented the first POR technique that is applicable to verifying trace equivalence properties of security protocols, without any action-determinism assumption. The key ingredients of this technique are: an equivalence LTS that recasts trace equivalence as a reachability property; a symbolic abstraction of the equivalence LTS on which persistent and sleep set techniques can be effectively computed; a collapse of conditionals that significantly speeds up these computations. Our technique applies to a wide class of protocols, has been implemented as a library and integrated in the state-of-the-art verifier DeepSec, showing significant performance improvements on case studies.

Compared to (our) earlier work on POR for protocol equivalences [8,9], we follow a radically different approach in this paper to obtain a technique that applies without any action-determinism assumption. In the action-deterministic case, the two techniques achieve similar but incomparable reductions: sleep sets are more efficient on *improper blocks*, but the focused behavior of *compression* is unmatched with sleep sets. Finally, we note that although sleep sets allow to recover a form of *dependency constraint*, we do not know how to justify its use in practice outside of the action-deterministic case. We hope that future work will allow to unify and generalize both techniques.

A crucial aspect of our new approach is that it manages to leverage classic POR techniques, namely persistent and sleep sets, for use in our specific security setting. In fact, we view this work as a first step towards bridging the gap between standard POR and security-specific techniques. As usual in POR, many variations (*e.g.*, in how we integrate with the equivalence verifiers) and approximations (*e.g.*, in independencies or stubborn set computations) should be explored to look for performance gains. The recent work on *dynamic POR* [25,4] (DPOR), which aims to find a trade-off between performance and quality of the computed persistent sets, is of particular interest here, though it is unclear at this point to which extent generic results can be extracted from the above-mentioned works for re-use in our security setting.

## A  Appendix of Section 2

In this section, we develop an example whose purpose is to show a protocol that can be modelled in our syntax, and how a privacy-type property, namely unlinkability, can be expressed through the notion of trace equivalence.

For illustrative purposes, we consider the BAC protocol used in e-passports. This protocol aims at establishing a fresh session key derived from $k_P$ and $k_R$. This session key will be used to encrypt the subsequent communication, and protect the information that are exchanged between the passport and the reader. Informally, we have:

1. $P \to R : n_P$
2. $R \to P : \{n_R, n_P, k_R\}_{ke}, \mathsf{mac}(\{n_R, n_P, k_R\}_{ke}, km)$
3. $P \to R : \{n_P, n_R, k_P\}_{ke}, \mathsf{mac}(\{n_P, n_R, k_P\}_{ke}, km)$

where $\{m\}_k$ denotes the symmetric encryption of a message $m$ with the key $k$, and mac is a keyed hash function. The keys $ke$ and $km$ are long term keys shared between the passport and the reader. First, $P$ sends a fresh random number $n_P$ to the reader, and the reader answers to this challenge by generating its own nonce $n_P$, as well as $k_R$ to contribute to fresh session key. This encryption together with a mac is sent to the passport. The passport will then check the mac, and decrypt the ciphertext and verify whether the nonce inside corresponds to the nonce $n_P$ generated at the first step. In case, decryption fails or the nonce inside the message is not the expected one, an error message will be sent. Otherwise, the message 3 is sent to the reader. After checking that the message is the expected one, both entities are able to compute the fresh session key derived form $k_R$ and $k_P$.

**Term algebra.** We consider the signature

$$\Sigma_{\mathsf{enc}} = \{\mathsf{enc}, \mathsf{dec}, \mathsf{mac}, \langle \ \rangle, \mathsf{proj}_1, \mathsf{proj}_2, \mathsf{nonce}_{\mathsf{err}}, \mathsf{mac}_{\mathsf{err}}\}.$$

The symbols enc, dec, and mac of arity 2 represent encryption and decryption, and message authentication code; concatenation of messages is modelled through the symbol $\langle \ \rangle$ of arity 2, with projection functions $\mathsf{proj}_1$ and $\mathsf{proj}_2$ of arity 1. The function symbols $\mathsf{nonce}_{\mathsf{err}}$ and $\mathsf{mac}_{\mathsf{err}}$ are constants (arity 0) that are used to model error messages. Then, we reflect the properties of the cryptographic primitives through the equational theory induced by the following equations:

$$\mathsf{dec}(\mathsf{enc}(x,y),y) = x, \ \mathsf{proj}_1(\langle x,y \rangle) = x, \text{ and } \mathsf{proj}_2(\langle x,y \rangle) = y.$$

**Processes.** The process $P(ke, km)$ modelling the role of the passport is given below. The process $R(ke, km)$ modelling the role of the reader can be written in a similar way.

$P(ke, km) = \mathsf{out}(c, n_P).\mathsf{in}(c, x).$
  if $\mathsf{mac}(\mathsf{proj}_1(x), km) = \mathsf{proj}_2(x)$ then
    if $\mathsf{proj}_1(\mathsf{proj}_2(\mathsf{dec}(\mathsf{proj}_1(x), ke))) = n_P$ then $\mathsf{out}(c, \langle m_P, \mathsf{mac}(m_P, km) \rangle)$
                                                                         else $\mathsf{out}(c, \mathsf{nonce}_{\mathsf{err}})$
                                       else $\mathsf{out}(c, \mathsf{mac}_{\mathsf{err}})$

where $m_P = \mathsf{enc}(\langle n_P, \langle \mathsf{proj}_1(\mathsf{dec}(\mathsf{proj}_1(x), ke)), k_P \rangle \rangle, ke)$.

Consider the two following configurations:

$$\mathcal{K}_{\mathsf{same}} = (P(ke, km); \phi_0), \text{ and } \mathcal{K}_{\mathsf{diff}} = (P(ke', km'); \phi_0)$$

with $\phi_0 = \{\mathsf{w}_{c',0} \triangleright \langle m'_R, \mathsf{mac}(m'_R, km) \rangle\}$ and $m'_R = \mathsf{enc}(\langle n'_R, \langle n'_P, k'_R \rangle \rangle, ke)$.

Intuitively, the configuration $\mathcal{K}_{\mathsf{same}}$ represents a situation where the passport in presence is the one for which we have already listen to an execution (the interesting message being stored in $\phi_0$), whereas $\mathcal{K}_{\mathsf{diff}}$ models the fact that the attacker is now in presence of an other passport. If $\mathcal{K}_{\mathsf{same}} \not\sqsubseteq_t \mathcal{K}_{\mathsf{diff}}$, this means that the attacker is able to perform a test that allows him to detect that the passport in presence is the same again.

Actually, we have that $\mathcal{K}_{\mathsf{same}} \xmapsto{\mathsf{tr}_0} (0; \phi)$ with
- $\mathsf{tr}_0 = \mathsf{out}(c, \mathsf{w}_{c,0}).\mathsf{in}(c, \mathsf{w}_{c',0}).\tau.\tau.\mathsf{out}(c, \mathsf{w}_{c,1})$, and
- $\phi = \phi_0 \uplus \{\mathsf{w}_{c,0} \triangleright n_P; \mathsf{w}_{c,1} \triangleright \mathsf{nonce}_{\mathsf{err}}\}$

It is possible to produce the same observable actions starting from $\mathcal{K}_{\mathsf{diff}}$, but this will result in the frame $\phi' = \phi_0 \uplus \{\mathsf{w}_{c,0} \triangleright n_P; \mathsf{w}_{c,1} \triangleright \mathsf{mac}_{\mathsf{err}}\}$. It is easy to see that $\phi \not\sim_s \phi'$. The test $\mathsf{w}_{c,1} = \mathsf{nonce}_{\mathsf{err}}$ holds in $\phi$ but not in $\phi'$, and thus $\mathcal{K}_{\mathsf{same}} \not\sqsubseteq \mathcal{K}_{\mathsf{diff}}$. This corresponds to a well-known replay attack discovered in [5] on French passports. This attack can be easily fixed by using the same error message in both cases. In such a case, the inclusion holds. This is a non trivial inclusion that can be automatically established by automatic verification tools such as DeepSec.

For illustrative purposes, we have only considered here a simple scenario for which configurations under study are actually action-deterministic according to the definition recalled below (and used in previous works).

**Definition 10.** *A configuration $A$ is* action-deterministic *if whenever $A \xmapsto{\mathsf{tr}} (\mathcal{P}; \Phi)$, and $P, Q$ are two distinct elements of $\mathcal{P}$, we have that $P$ and $Q$ cannot both perform an observable action of the same nature (*`in`*, *`out`*) on the same channel (i.e., if both actions are of same nature, their first argument has to differ).*

However, as soon as we want to consider more complex scenarios involving several passports and readers, we will have to deal with configurations that are not action-deterministic. Indeed, several passports will be able to output a message on channel $c$ at the same time. Using different channels (*e.g.*, one for each process in parallel) as it is often done (in so-called *simple* processes) leads to an imprecise modelling, and false attacks. Consequently, strong unlinkability is expressed as an equivalence between processes that are not action-deterministic [5].

When considering strong unlinkability, we also note that using *diff-equivalence* instead of trace equivalence, as is done in Tamarin and ProVerif when checking equivalences for unbounded sessions, systematically leads to false attacks [27]. Therefore, proposing POR techniques that are able to deal with non-action-deterministic processes is an important research goal. Actually, as far as we know, the only tool which is able to analyse the unlinkability property as described above on the BAC protocol is the tool UKANO [27]: it deals with an unbounded number of sessions but only supports a very specific class of 2-party protocols (and, of course, it is not a decision procedure). Otherwise, the only generic verification tool that is able to decide such a property is DeepSec and, like many other tools dedicated to protocols verification, it suffers from the usual state explosion problem for which we propose a solution in this paper.

# B Proofs of Section 3

This section contains proofs originally presented in [26], adapted to our style of presentation.

Two sequences of actions

$$w = \alpha_1 \ldots \alpha_i \alpha \beta \alpha_{i+3} \ldots \alpha_n \text{ and } w' = \alpha_1 \ldots \alpha_i \beta \alpha \alpha_{i+3} \ldots \alpha_n$$

that are executable from $s_0$ (*i.e.*, $s_0 \xrightarrow{w} s$ and $s_0 \xrightarrow{w'} s'$ for some $s$ and $s'$) are *equivalent* when $\alpha$ and $\beta$ are independent in $s_i$, where $s_i$ is the unique state such that $s_0 \xrightarrow{\alpha_1 \ldots \alpha_i} s_i$. The class of a trace starting from $s_0$ according to this equivalence relation is written $[w]_{s_0}$.

## B.1 Persistent sets

**Lemma 1.** *Let $s$ and $s'$ be two states such that $s \xrightarrow{w} s' \not\rightarrow$ with a non-empty sequence $w$ of actions. For any $s_0$ and $w_0$ such that $s_0 \xrightarrow{w_0} s$, there exists $\alpha \in \mathsf{p_{set}}(s_0, w_0)$ and $w'$ such that $\alpha w' \in [w]_s$.*

*Proof.* Let $w = \alpha_1 \alpha_2 \ldots \alpha_n$, and consider the following execution trace in the original LTS

$$s \xrightarrow{\alpha_1} s_1 \xrightarrow{\alpha_2} s_2 \ldots \xrightarrow{\alpha_{n-1}} s_{n-1} \xrightarrow{\alpha_n} s_n = s' \not\rightarrow$$

We show that, for some $1 \le i \le n$, $\alpha_i \in \mathsf{p_{set}}(s_0, w_0)$. We reason by contradiction, assuming the contrary. By definition of a persistent set, we then know that each $\alpha_i$ is independent in $s_{i-1}$ with all the actions in $\mathsf{p_{set}}(s_0, w_0)$. Therefore, we have that all actions in the non-empty set $\mathsf{p_{set}}(s_0, w_0)$ remain enabled in states $s_j$ with $0 < j \le n$, and thus they are enabled in $s'$ which is absurd.

Let $\alpha_{i_0}$ be the first action in $w$ that is in $\mathsf{p_{set}}(s_0, w_0)$. By definition of a persistent set, we know that for all $1 \le j < i_0$, we have $\alpha_{i_0}$ and $\alpha_j$ are independent in $s_{j-1}$. Therefore, considering the sequence $w' = \alpha_{i_0} \alpha_1 \ldots \alpha_{i_0-1} \alpha_{i_0+1} \ldots \alpha_n$, *i.e.*, the sequence $w$ in which the action $\alpha_{i_0}$ has been moved at the first position, we have that $w' \in [w]_s$. □

**Lemma 2** ([26]). *Let $s$ and $s'$ be two states such that $s \xrightarrow{w_1} s'$ (in the original LTS) for some sequence $w_1$. We have that $s \xrightarrow{w_2} s'$ for any $w_2 \in [w_1]_s$.*

*Proof.* By definition, all $w' \in [w]_s$ can be obtained from $w$ by successively permuting pairs of adjacent actions. It is thus sufficient to prove that, for any two sequences $w_1$ and $w_2$ that differ only by the order of two adjacent actions that are independent w.r.t. the current state, if $s \xrightarrow{w_1} s'$ then $s \xrightarrow{w_2} s'$. We assume that $w_1 = \alpha_1 \ldots \alpha \beta \ldots \alpha_n$ and $w_2 = \alpha_1 \ldots \beta \alpha \ldots \alpha_n$. By hypothesis, we have that:

$$s \xrightarrow{\alpha_1} s_1 \xrightarrow{\alpha_2} s_2 \ldots \xrightarrow{\alpha_i} s_i \xrightarrow{\alpha} s_{i+1} \xrightarrow{\beta} s_{i+2} \ldots \xrightarrow{\alpha_n} s_n = s'.$$

Since $\alpha$ and $\beta$ are independent in $s_i$, we have that:

$$s \xrightarrow{\alpha_1} s_1 \xrightarrow{\alpha_2} s_2 \ldots \xrightarrow{\alpha_i} s_i \xrightarrow{\beta} s'_{i+1} \xrightarrow{\alpha} s'_{i+2}.$$

Moreover, we know that $s_{i+2} = s'_{i+2}$. Therefore, we obtain that $s \xrightarrow{\alpha_1} s_1 \xrightarrow{\alpha_2} s_2 \ldots \xrightarrow{\alpha_i} s_i \xrightarrow{\beta} s'_{i+1} \xrightarrow{\alpha} s_{i+2} \ldots \xrightarrow{\alpha_n} s_n = s'$. This allows us to conclude. □

**Proposition 1.** *Let $s'$ be a final state that is reachable from $s$. We have that $s'$ is also reachable from $s$ through a trace that is persistent.*

*Proof.* Consider a trace $s = s_0 \xrightarrow{\alpha_1} s_1 \xrightarrow{\alpha_2} \ldots \xrightarrow{\alpha_n} s_n = s' \not\rightarrow$ that is not persistent. Let $k$ be the smallest index such that $\alpha_k \notin \mathsf{p}_{\mathsf{set}}(s_0, \alpha_1 \ldots \alpha_{k-1})$. Such a $k$ exists, and we have that $1 \leq k \leq n$. We proceed by induction on $n - k$.

The trace $w = \alpha_k \ldots \alpha_n$ is non-empty and executable in $s_{k-1}$. By Lemma 1, there is $\alpha w' \in [w]_{s_{k-1}}$ such that $\alpha \in \mathsf{p}_{\mathsf{set}}(s_0, \alpha_1 \ldots \alpha_{k-1})$. By Lemma 2, the trace $\alpha_1 \ldots \alpha_{k-1} \alpha w'$ can be executed from $s_0$ to $s_n$. It has the same length as our initial trace and it is either persistent or has a larger first non-persistent index. We conclude in both cases, either directly or by induction hypothesis. □

### B.2 Stubborn sets

In practice, persistent sets can be computed as the enabled actions of a *conditional stubborn set*, as defined and justified next.

**Definition 11.** *A set of actions $T$ is a* conditional stubborn set *for a state $s$ if $T \cap E(s) \neq \emptyset$ and, for all $\alpha \in T$, for all $s = s_1 \xrightarrow{\alpha_1} \ldots s_n \xrightarrow{\alpha_n}$ such that $\alpha \not\leftrightarrow_{s_n} \alpha_n$, we have $\alpha_i \in T$ for some $1 \leq i \leq n$.*

**Proposition 9.** *If $T$ is a conditional stubborn set for $s$, then $T \cap E(s)$ is a persistent set for $s$.*

*Proof.* Consider a trace $s = s_1 \xrightarrow{\alpha_1} s_2 \xrightarrow{\alpha_2} \ldots \xrightarrow{\alpha_{n-1}} s_n \xrightarrow{\alpha_n} s_{n+1}$ such that $\alpha_i \notin T \cap E(s)$ for any $i \in [1; n]$.

Let us first establish that $\alpha_i \notin T$ for any $i \in [1; n]$. By contradiction, consider the smallest $i$ such that $\alpha_i \in T$. Because $\alpha_i \notin T \cap E(s)$, this means that $\alpha_i$ is disabled in $s$ but enabled in $s_i$. Hence there is a $j < i$ such that $\alpha_j \not\leftrightarrow_{s_j} \alpha_i$. By definition of conditional stubborn sets, we conclude that there is a $k \leq j$ such that $\alpha_k \in T$, which contradicts the minimality of $i$.

We can now prove that $\alpha_n$ is independent in $s_n$ with any action $\alpha \in T \cap E(s)$. Assume the contrary, i.e., $\alpha \not\leftrightarrow_{s_n} \alpha_n$. By definition of conditional stubborn sets, we would then have $\alpha_i \in T$ for some $1 \leq i \leq n$, contradicting our first observation. □

### B.3 Sleep sets

**Proposition 2.** *If a final state is reachable from $s$ in the original LTS, then it is also reachable from $(s, \emptyset)$ through a sleep set execution.*

*Proof.* By Proposition 1 it is sufficient to show that any persistent execution towards a final state can be mapped to a sleep set execution. We consider a persistent execution $s_0 \xrightarrow{\alpha_1}_{\mathsf{p}_{\mathsf{set}}} s_1 \ldots \xrightarrow{\alpha_n}_{\mathsf{p}_{\mathsf{set}}} s_n \not\rightarrow$ with $s_n$ final, and set out to turn it into a sleep set execution from $(s_0, \emptyset)$ to some $(s_n, z_n)$.

By well-foundedness of the lexicographic ordering over $T^n$, we can assume that $w = \alpha_1 \ldots \alpha_n$ is minimal among the traces $w' \in [w]_{s_0}$ that are persistent.

Given one such execution, it suffices to show that, if there exists $p < n$ and $z_0, z_1, \ldots z_p$ such that $z_0 = \emptyset$, $(s_{i-1}, z_{i-1}) \xrightarrow{\alpha_i} (s_i, z_i)$ for all $1 \leq i \leq p$, then there exists $z_{p+1}$ such that $(s_p, z_p) \xrightarrow{\alpha_{p+1}} (s_{p+1}, z_{p+1})$. In fact, it suffices to show that $\alpha_{p+1}$ can be executed in $(s_p, z_p)$, and the existence of $z_{p+1}$ will follow.

Assume, by contradiction, that $\alpha_{p+1}$ cannot be executed. Since $\alpha_{p+1} \in \mathsf{p_{set}}(s_0, \alpha_1 \ldots \alpha_p)$ it must be the case that $\alpha_{p+1} \in z_p$. Let $k < p$ be the index at which $\alpha_{p+1}$ has last entered the sleep set, i.e., $\alpha_{p+1} \in z_i$ for all $i \in [k+1; p]$ but $\alpha_{p+1} \notin z_k$. We have:

- $\alpha_{p+1} \in \mathsf{p_{set}}(s_0, \alpha_1 \ldots \alpha_k)$ and $\alpha_{p+1} < \alpha_{k+1}$;
- $\alpha_{p+1} \leftrightarrow_{s_{i-1}} \alpha_i$ for all $i \in [k+1; p]$.

Consider the trace $w' = \alpha_1 \ldots \alpha_k \alpha_{p+1} \alpha_{k+1} \ldots \alpha_n$ obtained by executing $\alpha_{p+1}$ just after $\alpha_k$. We have $w' \in [w]_{s_0}$, but also $w' <_{\mathsf{lex}} w$. By Lemma 2, $w'$ executes from $s_0$ to $s_n$. It is not necessarily the case that $w'$ is a persistent trace since the choice of persistent sets may be affected after $\alpha_{p+1}$. However, we can apply $Proposition\ 1$ to $w'$ since $s_n$ is final. A simple inspection of the proof of Proposition 1 shows that we will more precisely obtain a $w'' \in [w']_{s_0} = [w]_{s_0}$, which will share at least the persistent prefix $\alpha_1 \ldots \alpha_k \alpha_{p+1}$ with $w'$. Hence $w'' <_{\mathsf{lex}} w$, which contradicts the minimality of $w$ among the persistent traces of $[w]_{s_0}$. □

## C Proof of Proposition 3

**Proposition 3.** *Let $A_0$ and $B_0$ be two alive configurations of same domain, and $s_0 = \langle|\mathbb{A}_0 \approx \mathbb{B}_0|\rangle$ where $\mathbb{A}_0 = \{A \mid A_0 \xmapsto{\tau^*} A \not\xmapsto{\tau}\}$, and $\mathbb{B}_0 = \{B \mid B_0 \xmapsto{\tau^*} B \not\xmapsto{\tau}\}$. The following conditions are equivalent:*

1. *$A_0$ is trace included in $B_0$, i.e., $A_0 \sqsubseteq_t B_0$;*
2. *no left-bad state is reachable from $s_0$ in the trace-equivalence LTS;*
3. *no left-bad, final state is reachable from $s_0$ in the trace-equivalence LTS.*

Before proving this proposition, we introduce three lemmas which state basic properties of the transition relation of our trace equivalence LTS.

**Lemma 3.** *Let $s = \langle|\mathbb{A} \approx \mathbb{B}|\rangle$ and $s' = \langle|\mathbb{A}' \approx \mathbb{B}'|\rangle$ be two states such that $s \xrightarrow{\mathsf{tr}} s'$ for some $\mathsf{tr}$. Let $(\mathcal{P}'; \phi') \in \mathbb{A}'$ (resp. $\mathbb{B}'$) be an alive configuration. In such a case, there exists $(\mathcal{P}; \phi) \in \mathbb{A}$ (resp. $\mathbb{B}$) such that $(\mathcal{P}; \phi) \xmapsto{\mathsf{tr}'} (\mathcal{P}'; \phi')$ for some $\mathsf{tr}'$ such that $\mathsf{obs}(\mathsf{tr}') = \mathsf{tr}$.*

**Lemma 4.** *Let $(\mathcal{P}; \phi)$, $(\mathcal{P}'; \phi')$ be two quiescent configurations such that $(\mathcal{P}; \phi) \xmapsto{\mathsf{tr}} (\mathcal{P}'; \phi')$ for some $\mathsf{tr}$. Let $s = \langle|\mathbb{A} \approx \mathbb{B}|\rangle$ be a state such that $(\mathcal{P}; \phi) \in \mathbb{A}$ (resp. $\mathbb{B}$). In such case, there exists $s' = \langle|\mathbb{A}' \approx \mathbb{B}'|\rangle$ such that $s \xrightarrow{\mathsf{obs}(\mathsf{tr})} s'$, and $(\mathcal{P}'; \phi') \in \mathbb{A}'$ (resp. $(\mathcal{P}'; \phi') \in \mathbb{B}'$).*

**Lemma 5.** *Let $W \subseteq \mathcal{W}$, $s = \langle|\mathbb{A} \approx \mathbb{B}|\rangle$ with domain $W$, and $s' = \langle|\mathbb{A}' \approx \mathbb{B}'|\rangle$ be a state such that $s \xrightarrow{\mathsf{tr}} s'$ for some $\mathsf{tr}$. We have that:*

- *for all $j$ and $(\mathcal{P}'; \phi') \in \mathbb{B}'^{\geq j}$, there exists $(\mathcal{P}; \phi'|_W) \in \mathbb{B}^{\geq j}$;*

– for all $j$ and $(\mathcal{P};\phi) \in \mathbb{B}^{\geq j}$, there exists $(\mathcal{P}';\phi') \in \mathbb{B}'^{\geq j}$ such that $\phi'|_W = \phi$.
*A similar result holds for $\mathbb{A}$.*

We are now ready to prove Proposition 3

*Proof (Proposition 3).* We first establish that $\neg(1) \Rightarrow \neg(2)$. We have $A_0 \overset{\text{tr}}{\mapsto} (\mathcal{P};\phi)$ for some $\mathcal{P}$, $\phi$ and $\text{tr}$ such that, for any $\mathcal{Q}$, $\psi$ and $\text{tr}'$ such that $B_0 \overset{\text{tr}'}{\mapsto} (\mathcal{Q};\psi)$ and $\text{obs}(\text{tr}) = \text{obs}(\text{tr}')$, we have $\phi \not\sim_s \psi$.

We can assume w.l.o.g. that $(\mathcal{P};\phi)$ is quiescent. The trace $\text{tr}$ can be written as $\text{tr}_1\text{tr}_2$ where $\text{tr}_2$ starts with an observable action and $\text{tr}_1$ contains only $\tau$ actions. Let $A$ be the quiescent configuration such that $A_0 \overset{\text{tr}_1}{\mapsto} A \overset{\text{tr}_2}{\mapsto} (\mathcal{P};\phi)$.

Thanks to Lemma 4, we have that $s_0 \overset{\text{tr}}{\to} s$ for some $s = \langle|\mathbb{A} \approx \mathbb{B}|\rangle$ such that $(\mathcal{P};\phi) \in \mathbb{A}$. We now assume, by contradiction, that $s$ is not a left-bad state. We have that $(\mathcal{P};\phi)$ is an alive configuration, and since we have assumed that $s$ is not left-bad, there exists $(\mathcal{Q};\psi) \in \mathbb{B}^{\geq \text{age}(s)}$ such that $\phi \sqsubseteq_s \psi$. First, we may note that $\mathbb{B}^{\geq \text{age}(s)}$ only contains alive configurations. Thus, we have that $\phi \sim_s \psi$ and thanks to Lemma 3, we know that $B \overset{\text{tr}'}{\mapsto} (\mathcal{Q};\psi)$ for some $\text{tr}'$ such that $\text{obs}(\text{tr}') = \text{obs}(\text{tr})$, which is a contradiction.

Now, we show that $\neg(2)$ implies $\neg(3)$. We assume that there is bad state $s = \langle|\mathbb{A} \approx \mathbb{B}|\rangle$ that is reachable from $s_0$. Let $n = \text{age}(s)$, $W$ the domain of $s$, and $\text{tr}$ be such that $s_0 \overset{\text{tr}}{\to} s$. Let $s' = \langle|\mathbb{A}' \approx \mathbb{B}'|\rangle$ be a final state such that $s \overset{\text{tr}'}{\to} s'$ for some $\text{tr}'$. Since $s$ is left-bad, we know that there exists $(\mathcal{P};\phi) \in \mathbb{A}$ such that:
– either $(\mathcal{P};\phi)$ is dead, *i.e.*, $\mathcal{P} = \bot_j$ with $0 < j < n$, and $\phi \not\sqsubseteq_s \mathbb{B}^{\geq j}$;
– or $(\mathcal{P};\phi)$ is alive and $\phi \not\sqsubseteq_s \mathbb{B}^{\geq n}$.
Our goal is to establish that $s'$ is a left-bad state too.

In case $(\mathcal{P};\phi)$ is dead with $\mathcal{P} = \bot_j$, we have that $(\mathcal{P};\phi) \in \mathbb{A}'$. We assume that $\phi \sqsubseteq_s \mathbb{B}'^{\geq j}$ and we show that this leads us to a contradiction. Indeed, if $\phi \sqsubseteq_s \psi'$ for some $\psi' \in \mathbb{B}'^{\geq j}$, then we have that $\phi \sqsubseteq_s \psi'|_W$ (since $\text{dom}(\phi) \subseteq W$). Now, thanks to Lemma 5, we get that $\psi'|_W \in \phi(\mathbb{B}^{\geq j})$, which contradicts $\phi \not\sqsubseteq_s \mathbb{B}^{\geq j}$.

In case $(\mathcal{P};\phi)$ is alive, thanks to Lemma 5, we know that there exists $A' = (\mathcal{Q};\phi') \in \mathbb{A}'^{\geq n}$ such that $\phi'|_W = \phi$. Independently of whether $A'$ is dead or alive, to conclude, it is sufficient to establish that $\phi' \not\sqsubseteq_s \mathbb{B}'^{\geq n}$. If $\phi' \sqsubseteq_s \psi'$ for some $\psi' \in \mathbb{B}'^{\geq n}$, then we have also that $\phi'|_W \sqsubseteq_s \psi'|_W$ for some $\psi' \in \mathbb{B}'^{\geq n}$. Thus, thanks to Lemma 5, we deduce that $\phi \sqsubseteq_s \mathbb{B}^{\geq n}$ leading to a contradiction.

To conclude, we show that $\neg(2)$ implies $\neg(1)$. Assume that $s_0 \overset{\text{tr}'}{\to} s' = \langle|\mathbb{A}' \approx \mathbb{B}'|\rangle$ for some $\text{tr}'$, such that $s'$ is left-bad and that $A_0$ is trace included in $B_0$. We seek for a contradicton. Let $W' = \text{dom}(s')$ and $n' = \text{age}(s')$. Moreover, we consider such a left-bad state for which $|\text{tr}'|$ is minimal in length. Since $s'$ is left-bad, we know that there exists $(\mathcal{P}';\phi') \in \mathbb{A}'$ such that:
– either $(\mathcal{P}';\phi')$ is dead, *i.e.*, $\mathcal{P}' = \bot_j$ with $0 < j < n'$, and $\phi' \not\sqsubseteq_s \mathbb{B}'^{\geq j}$;
– or $(\mathcal{P}';\phi')$ is alive and $\phi' \not\sqsubseteq_s \mathbb{B}'^{\geq n'}$.

Let $\text{tr}' = \text{tr}.\alpha$ and $s = \langle|\mathbb{A} \approx \mathbb{B}|\rangle$ (a configuration built on $W \subseteq W'$) such that $s_0 \overset{\text{tr}}{\to} s \overset{\alpha}{\to} s'$. Let $n = \text{age}(s)$, we have that $n \leq n'$, and we know that $s$ is *not* a left-bad state.

Assume $s'$ is bad due to the first item and $\mathcal{P}' = \bot_j$, then there is $(\mathcal{P}; \phi') \in \mathbb{A}$ (dead with $\mathcal{P} = \bot_j$; or alive and $j = n$), and since $s$ is not left-bad, we know that $\phi' \sqsubseteq_s \mathbb{B}^{\geq j}$. In any case, we have that $j \in \{0, \ldots, n\}$, and thanks to Lemma 5, we know that for any $\psi \in \mathbb{B}^{\geq j}$ there exists $\psi' \in \mathbb{B}'^{\geq j}$ such that $\psi'|_W = \psi$. Therefore, we deduce that $\phi' \sqsubseteq_s \mathbb{B}'^{\geq j}$ leading to a contradiction.

In case $(\mathcal{P}'; \phi')$ is alive and $\phi' \not\sqsubseteq_s \mathbb{B}'^{\geq n'}$. Thanks to Lemma 3, we know that $A \stackrel{\mathsf{tr}'}{\longmapsto} (\mathcal{P}'; \phi')$. We are going to show that for any $\psi'$ such that $B_0 \stackrel{\mathsf{tr}'_2}{\longmapsto} (\mathcal{Q}'; \psi')$ with $\mathsf{obs}(\mathsf{tr}'_2) = \mathsf{obs}(\mathsf{tr}')$, we have $\phi' \not\sim_s \psi'$. Assume, by contradcition, that there exists such a $(\mathcal{Q}'; \psi')$ with $\phi' \sim_s \psi'$. We can assume w.l.o.g. that $(\mathcal{Q}'; \psi'))$ is quiescent. In such a case, thanks to Lemma 4, we know that $(\mathcal{Q}'; \psi') \in \mathbb{B}'$, and because this is an alive configuration, we have $(\mathcal{Q}'; \psi') \in \mathbb{B}'^{\geq n'}$. This contradicts our hypothesis $\phi' \not\sqsubseteq_s \mathbb{B}'^{\geq n'}$. □

## D  Proofs of Section 5

We first mention a weak form of soundness that our symbolic semantics enjoys, and which will be useful in our proofs.

**Proposition 10 (soundness).** *Let $S$ and $S'$ be two symbolic states, $\theta \in \mathsf{Sol}(S)$, and $A$ be a symbolic action such that $S \stackrel{A}{\twoheadrightarrow} S'$.*
- *If $A = \mathtt{out}(c, \mathsf{w}_{c,i})$, then $\mathtt{out}(c, \mathsf{w}_{c,i}) \in E(S\theta)$.*
- *If $A = \mathtt{in}(c, \mathsf{X}^{c,i}, W)$, then $\mathtt{in}(c, R) \in E(S\theta)$ for any $R \in \mathcal{T}(\Sigma_0 \uplus \mathsf{dom}(S))$.*

*Proof.* The result directly follows from Definition 6. If $S \stackrel{A}{\twoheadrightarrow} S'$ then $S\theta$ can surely perform any plausible concretization of $A$: $A$ itself when it is an output or, when it is an input, any of its concretizations with a recipe that is valid wrt. $S\theta$. □

Note that, if $\alpha$ is the enabled concretization obtained above, there is no reason that $S\theta \stackrel{\alpha}{\to} S'\theta'$ for some $\theta'$: indeed, the branches of conditionals followed in the symbolic semantics to obtain $S'$ may be incompatible with $\theta$, and thus different from the branches followed in the concrete $\alpha$ transition. This is not a problem for the weak form of soundness that we require here.

**Proposition 4.** *Let $S = \langle \mathbb{A} \approx \mathbb{B} \rangle_{\mathcal{C}}^{I}$ be a symbolic state, $\theta \in \mathsf{Sol}(S)$. Let $s'$ and $\alpha$ be such that $S\theta \stackrel{\alpha}{\to} s'$. There exists $S'$, $A$ and $\theta' \sqsupseteq \theta$ such that $S \stackrel{A}{\twoheadrightarrow} S'$, $\theta' \in \mathsf{Sol}(S')$, $\alpha = A\theta'$, and $s' = S'\theta'$. Moreover, if $\alpha$ is of the form $\mathtt{in}(c, R)$, the proposition holds with $A = \mathtt{in}(c, \mathsf{X}^{c, I(c)}, W)$ for any $W$ such that $vars(R) \subseteq W$.*

*Proof.* The concrete transition $S\theta \stackrel{\alpha}{\to} s'$ consists of the execution of the visible action followed by some invisible steps, breaking parallels, evaluating conditionals, and performing choices by duplicating configurations. If $\alpha$ is an output, take $A = \alpha$; if $\alpha = \mathtt{in}(c, R)$, take $A = \mathtt{in}(c, \mathsf{X}^{c,i}, W)$ with $i = I(c)$ and an arbitrary $W \supseteq vars(R)$. There is a unique symbolic transition $S \stackrel{A}{\twoheadrightarrow} S'$ that performs $A$ and selects the same branches of conditionals, and this allows us to conclude. Indeed, the treatment of parallel compositions and choice operators is deterministic.

Define $\theta'$ to be $\theta$ when $\alpha$ is an output, and $\theta \uplus \{\mathsf{X}^{c,i} \mapsto R\}$ when $A = \mathtt{in}(c, \mathsf{X}^{c,i}, W)$. We check that $s' = S'\theta'$: this is obvious if $\alpha$ is an output, since in that case $\lambda_{\theta'} =$ $\lambda_\theta$. For an input, $\lambda_{\theta'}$ extends $\lambda_\theta$ with an instantiation for the new input variables: $\lambda_{\theta'}(\mathsf{x}_\psi^{c,i}) = (\mathsf{X}^{c,i}\theta')(\psi\lambda_{\theta'}) = R(\psi\lambda_{\theta'}) = R(\psi\lambda_\theta)$, which is the message that has been used as input in the concrete semantics for any configuration whose frame extends $\psi\lambda_\theta$.

We finally check that $\theta' \in \mathsf{Sol}(S')$. It has the correct domain to be a solution of $S'$. The domain of $\lambda_{\theta'}$ coincides with that of $\lambda_\theta$ for previous first-order variables. If any new variable $\mathsf{x}_\phi^{c,i}$ has appeared in $S'$, then it will be such that $\mathrm{dom}(\phi) = \mathrm{dom}(S) \cap W$, hence $\mathsf{x}_\phi^{c,i} \in \mathrm{dom}(\lambda_{\theta'})$. To conclude that $\theta' \in \mathsf{Sol}(S')$ it only remains to check that it satisfies the constraints of $S'$: it must be the case since $\lambda_\theta$ satisfied the constraints of $S$, and the newly added constraints correspond to conditions that held for $S\theta$. □

### D.1 Independencies

**Proposition 11.** *Let $S$ be symbolic state, and $A$ and $B$ be two symbolic actions executable in $S$ such that $A \Leftrightarrow_S^{ee} B$. Let $\theta \in \mathsf{Sol}(S)$, $s = S\theta$ and $\alpha$ (resp. $\beta$) be a concretization of $A$ (resp. $B$). If $\alpha, \beta \in E(s)$ then we have that $\alpha \leftrightarrow_s \beta$.*

*Proof.* We have that $\alpha$ and $\beta$ are enabled in $s$. We consider the respective successors $s_\alpha$ and $s_\beta$ of $s$ by these actions, and set out to show that $\alpha \in E(s_\beta)$ and $\beta \in E(s_\alpha)$, and $s$ has the same successor by $\alpha\beta$ and $\beta\alpha$.

By Proposition 4 we have that there exists $S_A$, $A'$, $\theta_A \sqsupseteq \theta$ such that $S \xrightarrow{A'} S_A$, $\theta_A \in \mathsf{Sol}(S_A)$, $\alpha = A'\theta_A$ and $s_\alpha = S_A\theta_A$, and similarly for $B$.

In case $\alpha$ is an output, since $\alpha$ is a concretization of both $A$ and $A'$ we conclude that $A = A'$. In case $\alpha$ is an input $\mathtt{in}(c, R)$, we know that $A = \mathtt{in}(c, X^{c,i}, W)$ for some $W$ such that $vars(R) \subseteq W$. Similarly, $A' = \mathtt{in}(c, \mathsf{X}^{c,i}, W')$ for some $vars(R) \subseteq W'$, but Proposition 4 tells us that we can choose $W' = W$, thus we can also assume that $A = A'$.

Therefore, we have that $S \xrightarrow{A} S_A$ and $S \xrightarrow{B} S_B$, and since $A \Leftrightarrow_S^{ee} B$, we know that $S_A \xrightarrow{B} S_{AB}$ and $S_B \xrightarrow{A} S_{BA}$ for some symbolic states $S_{AB}$ and $S_{BA}$. We will now rely on our Proposition 10. We have that $S_A \xrightarrow{B} S_{AB}$, and $\theta_A \in \mathsf{Sol}(S_A)$, thus we deduce that $s_\alpha \xrightarrow{\beta} s_{\alpha\beta}$ for some $s_{\alpha\beta}$. Similarly, we also have that $s_\beta \xrightarrow{\alpha} s_{\beta\alpha}$ for some $s_{\beta\alpha}$.

It now remains to establish that $s_{\alpha\beta} = s_{\beta\alpha}$. Applying Proposition 4 on $s_\alpha \xrightarrow{\beta} s_{\alpha\beta}$ we deduce that there exist $S_{AB}$, $B''$ and $\theta_{AB} \sqsupseteq \theta_A$ such that $S_A \xrightarrow{B''} S_{AB}$, $\theta_{AB} \in \mathsf{Sol}(S_{AB})$, $\beta = B''\theta_{AB}$, and $s_{\alpha\beta} = S_{AB}\theta_{AB}$. As before, we can assume that $B'' = B$. Similarly for $B$, we have that there exist $S_{BA}$ and $\theta_{BA} \sqsupseteq \theta_B$ such that $S_B \xrightarrow{A} S_{BA}$, $\theta_{BA} \in \mathsf{Sol}(S_{BA})$, $\alpha = A\theta_{BA}$, and $s_{\beta\alpha} = S_{BA}\theta_{BA}$.

We claim that $\theta_{AB} = \theta_{BA}$. Recall that the only variables in $\mathrm{dom}(\theta_{AB}) \smallsetminus \mathrm{dom}(\theta)$ are those introduced by $A$ (resp. $B$) in case $A$ (resp. $B$) is an input. Moreover, note that $A\theta_A = \alpha = A\theta_{BA}$ and $B\theta_B = \beta = B\theta_{AB}$. Hence, $\theta_A$ and $\theta_{BA}$ coincide on the variable potentially introduced by $A$. Since $\theta_A \sqsubseteq \theta_{AB}$, $\theta_{AB}$ and $\theta_{BA}$ coincide on the variable introduced by $A$. Similarly, they coincide on the variable introduced by $B$, hence they are equal.

Now, relying on the definition of $A \Leftrightarrow_S^{ee} B$, we deduce that $S_{AB}$ and $S_{BA}$ have incompatible constraints, or are equal. They cannot have incompatible constraints, since they share a solution $\theta_{AB} = \theta_{BA}$. Hence $S_{AB} = S_{BA}$, which allows us to conclude that $s_{\alpha\beta} = S_{AB}\theta_{AB} = S_{BA}\theta_{BA} = s_{\beta\alpha}$. □

**Proposition 12.** *Let $S$ be a symbolic state and $A$ and $B$ be two symbolic actions such that $A \Leftrightarrow_S^{de} B$. Let $\theta \in \mathsf{Sol}(S)$, and $s = S\theta$. Let $\alpha$ and $\beta$ be respective concretizations of $A$ and $B$. If $\alpha \notin E(s)$ and $\beta \in E(s)$ then $\alpha \leftrightarrow_s \beta$.*

*Proof.* Assume by contradiction that $\alpha \not\leftrightarrow_s \beta$, i.e., there is $s'$ and $s''$ such that $s \xrightarrow{\beta} s' \xrightarrow{\alpha} s''$. By Proposition 4, there exist $S'$, $S''$, $B'$, $A'$ and $\theta'' \sqsupseteq \theta$ such that $S \xrightarrow{B'} S' \xrightarrow{A'} S''$, $\theta'' \in \mathsf{Sol}(S'')$, $\alpha = A'\theta''$, $\beta = B'\theta''$, and $s'' = S''\theta''$.

If $\beta$ is an output then $B' = B$ is its only possible abstraction. If $\beta = \mathtt{in}(c, R)$, we have $B = \mathtt{in}(c, \mathsf{X}^{c,i}, W)$ with $vars(R) \subseteq W$, and $B' = \mathtt{in}(c, \mathsf{X}^{c,i}, W')$. However, Proposition 4 allows us to choose $W' = W$ in that case. Hence we can take $B' = B$ in both cases. The same reasoning on $\alpha$ allows us to take $A' = A$.

Therefore, we have that there exist $S'$ and $S''$ such that $S \xrightarrow{B} S' \xrightarrow{A} S''$. Relying on our hypothesis $A \Leftrightarrow_S^{de} B$, we deduce that $A$ is executable in $S$, and $A/B$ are not of the form $\mathtt{in}(c, \mathsf{X}^{c,i}, W)/\mathtt{out}(d, \mathsf{w}_{d,j})$ with $\mathsf{w}_{d,j} \in W$. Since $A$ is executable in $S$ but its concretization $\alpha \notin E(s)$, we deduce that $A$ is of the form $\mathtt{in}(c, \mathsf{X}^{c,i}, W)$, $\alpha$ is of the form $\mathtt{in}(c, R)$ with $vars(R) \subseteq W$ and $vars(R) \not\subseteq \mathrm{dom}(s) = \mathrm{dom}(S)$. Since $\alpha \in E(s')$, we know that $vars(R) \subseteq \mathrm{dom}(s') = \mathrm{dom}(S')$. To summarize, $\mathrm{dom}(S) \subsetneq \mathrm{dom}(R) \subseteq \mathrm{dom}(S')$. It must be that $B = \mathtt{out}(d, \mathsf{w}_{d,j})$ for some $d, j$ such that $\mathrm{dom}(S') = \mathrm{dom}(S) \uplus \{\mathsf{w}_{d,j}\}$, and $\mathsf{w}_{d,j} \in vars(R) \subseteq W$. This contradicts the fact that $A/B$ are not of the form $\mathtt{in}(c, \mathsf{X}^{c,i}, W)/\mathtt{out}(d, \mathsf{w}_{d,j})$ with $\mathsf{w}_{d,j} \in W$. □

**Proposition 5.** *Let $S$ be a symbolic state and $A$ and $B$ be two symbolic actions. Let $\theta \in \mathsf{Sol}(S)$, $s = S\theta$ and $\alpha$ (resp. $\beta$) be a concretisation of $A$ (resp. $B$).*
- *If $A \Leftrightarrow_S^{ee} B$, and $\alpha, \beta \in E(s)$, then $\alpha \leftrightarrow_s \beta$.*
- *If $A \Leftrightarrow_S^{de} B$, $\alpha \notin E(s)$ and $\beta \in E(s)$, then $\alpha \leftrightarrow_s \beta$.*

*Proof.* This is a direct consequence of Proposition 11 and Proposition 12. □

### D.2 Persistent set computation

**Proposition 6.** *Let $S$ be a symbolic state such that $EC(S) \neq \emptyset$, and consider $T = \{A\theta \mid A \in T^+(S)\}$. For any $\theta' \in \mathsf{Sol}(S)$, the set $T \cap E(S\theta')$ is persistent in $S\theta'$.*

*Proof.* By Proposition 9 it is sufficient to show that $T$ is stubborn for $S\theta'$ and contains at least one action enable in $S\theta'$. First, we check that $T$ contains at least one action enabled in $S\theta'$: indeed, we know that there exists $A \in T^+(S) \cap EC(S)$, and that it has a concretization enabled in $S\theta'$ by Proposition 10. Second, we consider $\alpha \in T$ and an execution
$$s = s_1 \xrightarrow{\alpha_1} \ldots s_n \xrightarrow{\alpha_n} s_{n+1}$$
such that $\alpha \not\leftrightarrow_{s_n} \alpha_n$, and shall establish that $\alpha_i \in T$ for some $1 \leq i \leq n$.

By Proposition 4, we know that there exist symbolic states $S_1, \ldots, S_{n+1}$, and symbolic actions $A_1, \ldots, A_n$, and substitutions $\theta_{n+1} \sqsupseteq \theta_n \ldots \sqsupseteq \theta_1 = \theta$ such that
$$S = S_1 \xrightarrow{A_1} S_2 \xrightarrow{A_2} \ldots \xrightarrow{A_{n-1}} S_n \xrightarrow{A_n} S_{n+1}$$

where $\theta_i \in \mathsf{Sol}(S_i)$, $s_i = S_i\theta_i$ for $1 \leq i \leq n+1$, and $\alpha_i = A_i\theta_{i+1}$ for $1 \leq i \leq n$. Moreover, we can take the $A_i$ to be in $EC(S_i)$.

By definition of $T$, we have that $\alpha$ is a concretization of some $A \in T^+(S)$. If there exists some $i \in [1;n]$ such that $A \not\Leftrightarrow_{S_i} A_i$, then the specification of $T^+(S)$ allows us to conclude that there exists $j \in [1;i]$ such that $A_j \in T^+(S)$. Thus $\alpha_j = A_j\theta_{j+1} \in T$, which concludes the argument in this case.

Assume now that $A \Leftrightarrow_{S_i} A_i$ for all $i \in [1;n]$. We distinguish two cases:
- *Case 1:* $\alpha \in E(s_1)$. By $\alpha \leftrightarrow_{s_1} \alpha_1$ we obtain that $\alpha \in E(s_2)$. Repeating this argument, we actually have $\alpha \in E(s_i)$ for all $1 \leq i \leq n$. Hence its abstraction $A$ is executable in $S_n$, as is the case for $A_n$. Thus $A \Leftrightarrow_{S_n} A_n$ implies $A \Leftrightarrow^{ee}_{S_n} A_n$ and by Proposition 11 we obtain that $\alpha \leftrightarrow_{s_n} \alpha_n$, which is a contradiction.
- *Case 2:* $\alpha \notin E(s_1)$. In such a case, since $\alpha \leftrightarrow_{s_1} \alpha_1$, we deduce that $\alpha \notin E(s_2)$, and eventually that $\alpha \notin E(s_n)$. Now, since we know that $A \Leftrightarrow_{S_n} A_n$, we deduce that $A \Leftrightarrow^{de}_{S_n} A_n$. We have a concretization $\alpha$ of $A$ that is not in $E(S_n\theta_n)$, and a concretization $\alpha_n$ of $A_n$ that is in $E(S_n\theta_n)$. By Proposition 12 we deduce that $\alpha \leftrightarrow_{s_n} \alpha_n$, which is again a contradiction. □

### D.3 Symbolic sleep sets

**Proposition 7.** *Let $(s_0, \emptyset) \xrightarrow{\alpha_0} (s_1, z_1) \ldots \xrightarrow{\alpha_{n-1}} (s_n, z_n)$ be a sleep set execution in our initial LTS. Then, there is $(S_0, \emptyset) \xrightarrow{A_0} (S_1, Z_1) \ldots \xrightarrow{A_{n-1}} (S_n, Z_n)$ a sleep set execution in our symbolic LTS, and substitutions $\emptyset = \theta_0 \sqsubseteq \theta_1 \ldots \sqsubseteq \theta_n$ such that $s_i = S_i\theta_i$, $\alpha_i = A_i\theta_{i+1}$ for $i \in [1; n-1]$, and $s_n = S_n\theta_n$.*

*Proof.* We will actually prove the result with $(S_0, \ldots, S_n) = \mathsf{abs}(s_0, \alpha_0, \ldots, \alpha_{n-1})$. Each $A_i$ is then uniquely determined as the only $A \in EC(S_i)$ which admits $\alpha_i$ as a concretization. We then proceed by induction on $n$, showing additionally that we have, for all $i \in [1;n]$, three additional invariants:

$(a)$ for all $A \in Z_i$, $A$ is executable in $S_i$
$(b)$ for all $\mathtt{in}(c, \mathsf{X}^{c,i}, W) \in Z_i, W \subseteq \mathsf{dom}(S_i)$
$(c)$ for all $A \in Z_i$, for all $\sigma$ such that $A\sigma \in E(S_i\theta_i)$, $A\sigma \in z_i$

Invariants $(a)$ and $(b)$ will immediately follow from the fact that we only put to sleep executable actions, and that they remain executable as long as we keep them in the sleep set (because they are $\Leftrightarrow^{ee}$ with the executed actions). Invariant $(c)$ states that all enabled concretizations of actions put to sleep in the symbolic semantics have been put to sleep in the concrete semantics.

If $n = 0$ the result is obviously satisfied with $S_0$ such that $S_0\theta_0 = s_0$ and $Z_0 = \emptyset$.

Assuming that the result holds for $n$, we will now establish it for $n+1$. The symbolic execution obtained from $\mathsf{abs}(s_0, \alpha_0, \ldots, \alpha_n)$ is an extension of the one obtained from $\mathsf{abs}(s_0, \alpha_0 \ldots \alpha_{n-1})$. By induction hypothesis we thus have some sleep sets such that $(S_{i-1}, Z_{i-1}) \xrightarrow{A_{i-1}} (S_i, Z_i)$ for all $i \in [1;n]$.

We first show that $A_n$ is executable in $(S_n, Z_n)$. By definition of $\mathsf{p_{set}}$, since $\alpha_n \in \mathsf{p_{set}}(s_0, \alpha_0 \ldots \alpha_{n-1})$, it is the concretization of an action in $T^+(S_n) \cap EC(S_n)$, which must be $A_n$ itself since $A_n \in EC(S_n)$ and $A_n\theta_{n+1} = \alpha_n$ (distinct actions in the enabled cover cannot share a concretization). It remains to check that $A_n \notin Z_n$. We assume the contrary and seek to derive a contradiction.

- If $A_n$ is an output, we have $A_n\sigma = \alpha_n$ for any $\sigma$. By Proposition 10, since $A_n$ is executable in $S_n$, $A_n\sigma \in E(S_n\theta_n)$, hence thanks to our induction hypothesis (item c) $A_n\sigma = \alpha_{n+1} \in z_{n+1}$, which contradicts the definition of concrete sleep set executions.
- If $A_n = \text{in}(c, \mathsf{X}^{c,i}, W)$, we have $W = \text{dom}(S_n)$ since $A_n \in EC(S_n)$. We also have $\alpha_n = \text{in}(c, R)$ and, since $\alpha_n \in E(s_n)$, $vars(R) \subseteq \text{dom}(s_n) = \text{dom}(S_n)$. Let $\sigma$ be such that $A_n\sigma = \alpha_n$. We have $\alpha_n = A_n\sigma \in E(S_n\theta_n)$ by Proposition 10. Thanks to our induction hypothesis (item c), we deduce that $\alpha_n \in z_n$ which is again a contradiction.

The value of $Z_{n+1}$ is forced by the (symbolic) sleep set semantics, as was the case for $z_{n+1}$. We only need to check our additional invariants for $Z_{n+1}$. All actions $A \in Z_{n+1}$ were executable in $S_n$ (either by (a) or by definition of the symbolic sleep transitions) and satisfy $A \Leftrightarrow^{ee}_{S_n} A_n$, hence they are still executable in $S_{n+1}$. Invariant (b) also holds, since $\text{dom}(S_n) \subseteq \text{dom}(S_{n+1})$, and we may only add inputs to $Z_{n+1}$ when their domain is $\text{dom}(S_n)$. We now check (c). Let $A \in Z_{n+1}$ and $\sigma$ be such that $A\sigma \in E(S_{n+1}\theta_{n+1})$. We show that $A\sigma \in z_{n+1}$, distinguishing two cases:

- We first consider the case where $A$ belonged to $Z_n$ and $A \Leftrightarrow^{ee}_{S_n} A_n$. Then, $A$ is executable in $S_n$ by (a), and by (b) the domain of $A$ (if it is an input) is included in $\text{dom}(S_n)$. Hence $A\sigma \in E(S_n\theta_n)$ by Proposition 10. We then have $A\sigma \in z_n$ by (c). By $A \Leftrightarrow^{ee}_{S_n} A_n$ and Proposition 11 we also have $A\sigma \leftrightarrow_{s_n} \alpha_n$, and hence $A\sigma \in z_{n+1}$ by definition of the concrete sleep set executions.
- Otherwise, $A$ has been added because $A \in T^+(S_n) \cap EC(S_n)$, $A < A_n$ and $A \Leftrightarrow^{ee}_{S_n} A_n$. We have the analogue in the concrete semantics: $A\sigma \in \mathsf{p}_{\mathsf{set}}(s_0, \alpha_0 \ldots \alpha_{n-1})$, $A\sigma < \alpha_n$ and finally $A\sigma \leftrightarrow_{s_n} \alpha_n$ by Proposition 11. Hence $A\sigma \in z_{n+1}$ by definition of the concrete sleep set executions. □

### D.4 Collapsing conditionals

The goal of this section is to prove the following result.

**Proposition 8.** *For any execution $S = S_0 \xrightarrow{A_1} S_1 \ldots \xrightarrow{A_n} S_n$, there is an execution $S^c = T_0 \xrightarrow{A_1} T_1 \ldots \xrightarrow{A_n} T_n$, such that, for any $A$ and $i \in [1; n]$, $A \Leftrightarrow_{T_{i-1}} A_i$ (resp. $A_i \Leftrightarrow_{T_{i-1}} A$) implies $A \Leftrightarrow_{S_{i-1}} A_i$ (resp. $A_i \Leftrightarrow_{S_{i-1}} A$).*

*Hence, $T^+(S^c)$ is a symbolic stubborn set for $S$ and any sleep set execution from $S$ is also a sleep set execution from $S^c$.*

We assume here that $S^c$ is obtained from $S$ by collapsing a single conditional, instead of applying repeatedly this process (see Section 5.5). The general result can be obtained by repeatedly applying the result limited to one collapse. Formally, we pose: $S = S'[T]$,
$$T = \big(\text{if } u = v \text{ then } \alpha.P_1 \text{ else } \beta.P_2\big),$$
$S^c = S'[T^c]$, $T^c = \gamma.\text{if } u = v \text{ then } P_1 \text{ else } P_2$ and:
- $\gamma = \text{in}(c, x)$ when $\alpha = \beta = \text{in}(c, x)$;
- $\gamma = \text{out}(c, \Delta(t_1, t_2, u, v))$ when $(\alpha, \beta) = (\text{out}(c, t_1), \text{out}(c, t_2))$.

Our plan is to explicit the states $T_1 \ldots T_n$ through an transformation $\pi$ applied uniformly on $S_1 \ldots S_n$: $T_i = \pi(S_i)$. Informally, we need to replace non-executed instances of $T$ by $T^c$. A different transformation should be applied to $T$ when it is evaluated but $\alpha$ and $\beta$ have not been executed. Once these actions are executed, $T$ and $T^c$ will be synchronized again (up to terms). However, there are two main pitfalls. First, even if $S$ and $S^c$ initially differ by only one conditional, this conditional can be duplicated along an execution in different configurations due to non-determinism. Worse, those different occurrences of the initial conditional that may occur in different configurations of a single $S_i$ may have evolved differently along the execution: because their surrounding frames may eventually differ, and thus their input terms may also differ. In order to keep track of all those occurrences of the initial conditional, we shall annotate the conditional, making them recognizable from the others. Second, we shall keep track of how potential variables of $u, v, t_1, t_2$ are instantiated by first-order variables as inputs preceding $T$ are executed through the execution. This is needed in order to know how to transform a potential occurrence a conditional in some $T_i$, typically for building $\Delta(t_1\rho, t_2\rho, u\rho, v\rho)$ with the appropriate substitution $\rho$.

*Annotations.* For the purpose of our proof, we equip the syntax of processes with optional annotations that have no impact on the semantics and are only useful for defining $\pi$. First, we formally explain how this is done for the case of outputs. When $(\alpha, \beta) = (\mathsf{out}(c, t_1), \mathsf{out}(c, t_2))$, $T$ shall be annotated as follows:

$$T = \mathtt{if}^* \; u =_\rho v \; \mathtt{then} \; \mathsf{out}^*(c, t_1[\rho]).P' \; \mathtt{else} \; \mathsf{out}^*(c, t_2[\rho]).Q'$$

for an initial substitution $\rho = \emptyset$. When $\alpha = \beta = \mathsf{in}(c, x)$, we shall use the following annotation:

$$T = \mathtt{if}^* \; u =_\rho v \; \mathtt{then} \; \mathsf{in}^*_\rho(c, x).P' \; \mathtt{else} \; \mathsf{in}^*_\rho(c, x).Q'$$

for an initial substitution $\rho = \emptyset$. The annotation $^*$ may thus label an action or a conditional construct, while the annotation $\rho$ keeps track of the instantiation from variables in $u, v$, (and $t_1$ and $t_2$ for the output case) to first-order variables. The semantics of processes is adapted straightforwardly to our annotated processes. In particular, substitutions performed upon input actions are propagated to $\rho$ annotations, where substitutions thus accumulate. Annotations being only a technical device to track information through an execution, and construct our transformed states $T_i$, it should of course be ignored, *e.g.*, when testing for independencies in a state $S_i$.

*The projection transformation $\pi$.* Given a state $S_i$, we define $\pi(S_i)$ by repeatedely applying the following transformations:
  – Any occurrence of

$$\mathtt{if}^* \; u' =_\rho v' \; \mathtt{then} \; \mathsf{out}^*(c, t'_1[\rho]).Q_1 \; \mathtt{else} \; \mathsf{out}^*(c, t'_2[\rho]).Q_2$$

is replaced by

$$\mathsf{out}(c, \Delta(t_1\rho, t_2\rho, u\rho, v\rho)).\mathtt{if} \; u\rho = v\rho \; \mathtt{then} \; Q_1 \; \mathtt{else} \; Q_2.$$

Note that the terms $u', v', t'_1, t'_2$ and the processes $Q_1, Q_2$ may differ from the terms $u, v, t_1, t_2$ and the processes $P_1, P_2$ that are initially in the conditional replaced in $S$ because of some variables being replaced by first-order variables in input transitions. However, we have $u' = u\rho$, $v' = v\rho$, $Q_1 = P_1\rho$, $Q_2 = P_2\rho$.
- Similarly, any occurrence of

$$\texttt{if}^*\ u' =_\rho v'\ \texttt{then}\ \mathsf{in}^*_\rho(c,x).Q_1\ \texttt{else}\ \mathsf{in}^*_\rho(c,x).Q_2$$

is replaced by

$$\mathsf{in}^*_\rho(c,x).\texttt{if}\ u\rho = v\rho\ \texttt{then}\ Q_1\ \texttt{else}\ Q_2.$$

- Any occurrence of $\mathsf{out}^*(c,t[\rho]).Q$ at top-level in a configuration is replaced by

$$\mathsf{out}(c, \Delta(t_1\rho, t_2\rho, u\rho, v\rho)).\texttt{if}\ u\rho = v\rho\ \texttt{then}\ P_1\rho\ \texttt{else}\ P_2\rho.$$

In this case, we will necessarily have $t = t_i\rho$ and $Q = P_i\rho$ for some $i \in \{1,2\}$.
- Any occurrence of $\mathsf{in}^*_\rho(c,x).Q$ at top-level in a configuration is replaced by

$$\mathsf{in}(c,x).\texttt{if}\ u\rho = v\rho\ \texttt{then}\ P_1\rho\ \texttt{else}\ P_2\rho.$$

- Any occurrence of $w \triangleright t[\rho]$ in a frame is replaced by $w \triangleright \Delta(t_1\rho, t_2\rho, u\rho, v\rho)$. Note that this includes frames of first-order variables, which themselves occur both in configurations and constraints.

In particular, note that $S^c = \pi(S)$.

For some symbolic state $T$ and a set of constraints $\mathcal{C}$, we note $T \wedge \mathcal{C}$ the state $T$ to which the constraints of $\mathcal{C}$ have been added. We now prove the following proposition, that trivially implies Proposition 8.

**Proposition 13.** *(a) For any $S = S_0 \xrightarrow{A_1} S_1 \ldots \xrightarrow{A_n} S_n$, there exist $\mathcal{C}_1 \ldots \mathcal{C}_n$ and an execution $S^c = T_0 \xrightarrow{A_1} T_1 \ldots \xrightarrow{A_n}$ such that $T_i \wedge \mathcal{C}_i = \pi(S_i)$ for all $1 \leq i \leq n$.*
*(b) Moreover, for any $A$ and $1 \leq i \leq n$, $A \Leftrightarrow_{T_{i-1}} A_i$ (resp. $A_i \Leftrightarrow_{T_{i-1}} A$) implies $A \Leftrightarrow_{S_{i-1}} A_i$ (resp. $A_i \Leftrightarrow_{S_{i-1}} A$).*
*(c) Hence, $T^+(S^c)$ is a symbolic stubborn set for $S$ and any sleep set execution from $S$ is also a sleep set execution from $S^c$.*

*Proof.* (a) This is obtained by induction on $n$, and we give below the key reasons why it holds. First note that applying $\pi$ never changes the set of actions executable in a state. More specifically, the states $T_i$ are obtained by making the same choices through conditionals as in $S_i$ (for the conditionals that are not postponed by $\pi$). Finally, we define the $\mathcal{C}_i$ to be the set of constraints added in $S_i$ because of conditionals modified by $\pi$ that have been performed in $S_i$ but not yet in $T_i$. Formally, $\mathcal{C}_i$ is the set of all $\pi(u\rho) = \pi(v\rho)$ (resp. $\pi(u\rho) \neq \pi(v\rho)$) such that an action carrying[4] the annotation $\rho$ is available at toplevel in a configuration of $S_i$, and $u\rho = v\rho$ (resp. $u\rho \neq v\rho$) is in the constraint of $S_i$ — the role of $\pi$ here is simply to introduce $\Delta$ symbols in frames occurring as part of first-order variables in $u$, $v$ and $\rho$.

(b) We actually prove the contrapositive, *i.e.*, no dependencies are introduced when applying $\pi$. This is easily checked for $\Leftrightarrow^{de}$ since $\pi$ does not change the executable actions, and since (a) guarantees that executions from $S_{i-1}$ can be mapped to executions from $T_{i-1}$ (by action-determism of the symbolic semantics). The first part of $\Leftrightarrow^{ee}$ holds for the same reasons, we thus concentrate on the second one. We consider some executions
$$S_{i-1} \xrightarrow{A_i} S_A \xrightarrow{B} S_{AB} \text{ and } S_{i-1} \xrightarrow{B} S_B \xrightarrow{A_i} S_{BA}$$
such that $S_{AB}$ and $S_{BA}$ have compatible constraints and $S_{AB} \neq S_{BA}$, and we seek to establish that there exist executions
$$T_i \xrightarrow{A_i} T_A \xrightarrow{B} T_{AB} \text{ and } T_i \xrightarrow{B} T_B \xrightarrow{A_i} T_{BA}$$
such that $T_{AB}$ and $T_{BA}$ have compatible constraints and $T_{AB} \neq T_{BA}$. This will imply $A_i \not\Leftrightarrow_{T_i} B$ as desired.

---

[4] In case of an input, $\rho$ is directly found in $\mathsf{in}^*_\rho(c,x)$. In case of an output $\mathsf{out}^*(c,t[\rho])$, $\rho$ is found in the term.

The existence of $T_{AB}$ and $T_{BA}$ is given by (a). We actually have $T_{AB} \wedge \mathcal{C}_{AB} = \pi(S_{AB})$, and similarly for other states.

Hence, all constraints of $T_{AB}$ (resp. $T_{BA}$) are in $\pi(S_{AB})$ (resp. $\pi(S_{BA})$). The only effect of $\pi$ on constraints is the last item of the transformation, *i.e.*, the introduction of $\Delta$ symbols in frames. Hence, if two incompatible constraints $C_1$ and $C_2$ existed in $T_{AB}$ and $T_{BA}$, they would be of the form $\pi(C_1')$ and $\pi(C_2')$ with $C_1'$ and $C_2'$ in $S_{AB}$ and $S_{BA}$. By inspecting the arguments of $\Delta$ terms introduced by $\pi$ in $\pi(C_1')$ and $\pi(C_2')$, one can easily deduce that $C_1'$ and $C_2'$ must be incompatible as well, contradicting our hypothesis.

We now verify that $T_{AB} \neq T_{BA}$. First, consider the case where $S_{AB} \neq S_{BA}$ because of a difference in constraints: there is a constraint $C$ in the constraints of $S_{AB}$ but not $S_{BA}$. This difference persists after the projection, *i.e.*, $\pi(S_{AB}) \neq \pi(S_{BA})$, because the effect of $\pi$ on constraints is only to expand some output messages into $\Delta$ terms, in an injective fashion ($\Delta$ terms contain all information from the original terms). Hence we have $T_{AB} \wedge \mathcal{C}_{AB} \neq T_{BA} \wedge \mathcal{C}_{BA}$. We immediately conclude if $\pi(C)$ is in $T_{AB}$. Otherwise, it is in $\mathcal{C}_{AB}$, which means that the conditional has been evaluated in $S_{AB}$ but not in $T_{AB}$, where it has been postponed after the collapsed action $\gamma$. The postponed conditional found in $T_{AB}$ will then be absent from any configuration of $T_{BA}$ having the same frame, which gives us $T_{AB} \neq T_{BA}$. Now assume that a configuration of $D$ of $S_{AB}$ is absent from $S_{BA}$. We show that $\pi(D)$ is absent from $T_{BA}$. It suffices to show that for all $D \neq D'$, we have $\pi(D) \neq \pi(D')$. If the mismatch is due to the skeleton of processes (*i.e.*, processes without terms) then it is preserved by $\pi$ and we conclude. Otherwise, as argued before, differences caused by terms on the $S$ side cannot disappear with $\pi$, because the introduction of $\Delta$ terms is injective:
$$\Delta(t_1^{AB}\rho, t_2^{AB}\rho, u^{AB}\rho, v^{AB}\rho) = \Delta(t_1^{BA}\rho, t_2^{BA}\rho, u^{BA}\rho, v^{BA}\rho)$$
implies both $t_1^{AB}\rho = t_1^{BA}\rho$ and $t_2^{AB}\rho = t_2^{BA}\rho$.

(c) We obtain that $T^+(S^c)$ is a symbolic stubborn set for $S$ as a consequence of (a) and (b). Finally, any sleep execution from $S$ built on top of $T^+(S^c)$ is an execution of $S^c$ by (a). Furthermore, (b) implies that sleep sets we would obtain from $S^c$ are necessarily smaller, which implies that such an execution is a correct sleep execution from $S^c$. □